\documentclass[prc]{revtex4-1}
\usepackage{graphicx}
\usepackage{amsmath}
\usepackage{multirow} 
\usepackage[nooneline]{caption}

\begin{document}

\title{Cooper pair transfer in nuclei}

\author{G. Potel$^{1}$}
\author{A. Idini$^{2,3,4}$}
\author{F. Barranco$^{5}$}
\author{E. Vigezzi$^{4}$}
\author{R.A. Broglia$^{3,4,6,7}$}

\affiliation{
$^{1}$
CEA-Saclay, IRFU/Service de Physique Nucl\'eaire, F-91191 Gif-sur-Yvette, France
\\
$^{2}$
Institut f\"ur Kernphysik, Technische Universit\"{a}t Darmstadt, Schlossgartenstrasse 2, 64289 Darmstadt, Germany
\\
$^{3}$ 
Dipartimento di Fisica, Universit$\grave{a}$ degli 
Studi di Milano, Milano, Italy
\\
$^{4}$ 
INFN, Sezione di Milano,  Milano, Italy
\\
$^{5}$
Departamento de Fisica Aplicada III, Escuela Superior de Ingenieros, Universidad de Sevilla, 
Sevilla, Spain
\\
$^{6}$
The Niels Bohr Institute, University of Copenhagen, Copenhagen, Denmark 
\\
$^{7}$
FoldLESs S.r.l. \\
via Valosa di Sopra 9, I--20052  Monza (MB), Italy
}

\date{\today}

\begin{abstract}
The second order DWBA implementation of two-particle transfer direct reactions
which includes simultaneous and successive transfer, properly corrected by non-orthogonality effects is tested with the help
of controlled nuclear structure and reaction inputs against data spanning the whole mass table, and showed to constitute  a quantitative 
probe of nuclear pairing correlations.
\end{abstract}

\maketitle

\section{Introduction}
Cooper pairs are the building blocks of pairing correlations in many-body fermionic systems \cite{Cooper:56}. In particular in atomic nuclei (see \cite{Bohr:58,Bohr:75,Brink:05} and refs. therein). As a consequence, nuclear superfluidity, similarly to condensed matter superconductivity, can be specifically and quantitatively probed through Cooper pair tunnelling (cf. \cite{Bohr:75,Yoshida:62,Bohr:64,Broglia:73b,Broglia:73a} and refs. therein). To do so in the nuclear case, one should be able to predict, making use of nuclear structure spectroscopic amplitudes, absolute differential cross sections, sole quantities which can be directly compared with the experimental observations.
In what follows we review some of the central developments which eventually made this requirement operative. Two technical remarks: a) throughout, light and heavy ions will be treated essentially on equal footing; b) for simplicity, only sphercal nuclei will be considered.

In the simultaneous transfer of two nucleons, one nucleon goes over from target to projectile, or viceversa, under the influence of the nuclear interaction
responsible of the existence of a mean field potential,  while the other follows suit by profiting of: 1) pairing correlations (simultaneous transfer); 
2) the fact that the single-particle wavefunctions describing the motion of Cooper pair partners in both target and projectile are solutions of different 
single-particle potentials (non-orthogonality transfer). 
In the limit of independent particle motion, in which all of the nucleon-nucleon interaction is assumed to be used up in generating a mean field, both contributions
to the transfer process (simultaneous and non-orthogonality) cancel out exactly (cf. App. A).

In keeping with the fact that nuclear Cooper pairs are weakly bound, this cancellation is, in actual nuclei, quite strong. Consequently, successive transfer, a process in which the mean field acts twice is, as a rule, the main mechanism at the basis of Cooper pair transfer (Fig. 1). Because of the same reason (weak binding), the correlation length of Cooper pairs is larger than nuclear dimensions, a fact which allows the two members of a Cooper pair, to move between target and projectile, essentially as a whole, also in the case of successive transfer. Within this context, it is of notice that, 
while one-particle transfer is a rather local process, Cooper pair tunneling is highly non-local (cf. Fig. 2).

The condensation of these extended and thus strongly overlapping, bosonic objects gives rise to a highly correlated, coherent, quasiclassical superfluid state, displaying overall phase coherence, and essentially exhausting two-nucleon transfer sum rules \cite{Broglia:72b}. Consequently, they are amenable to an accurate theoretical descriptions in terms of simple models, in particular that resulting from  the BCS approximation.

In the case of pairing correlations around closed shell nuclei, the associated pairing vibrational modes are also amenable to a quite accurate nuclear structure description in terms of the random phase approximation (RPA) for nuclei along the stability valley and of nuclear field theory (NFT) for light nuclei along the drip lines, taking into account, in both cases, ground state correlations, and coupling to the continuum, in particular in the case of halo nuclei. 

In other words, ground state correlations and medium polarization effects, together with the associated renormalization (self--energy and vertex correction phenomena) and induced interaction processes, are essential to obtain an accurate description of the nuclear structure in general, and of nuclear pairing phenomena in particular (cf. e.g. Chs. 5, 8, 9 and 10 of ref. \cite{Brink:05} and refs therein). Such  effects become the dominant ones in the case of light nuclei along the neutron drip lines, in which case the glue binding the halo nucleons to the core is essentially all of dynamic origin, continuum effects playing also an important role (cf. e.g. Ch. 11 of ref. \cite{Brink:05} and refs. therein).

With the help of the corresponding essentially "exact", two-nucleon transfer spectroscopic amplitudes associated with the superfluid state, and of empirical, global optical potentials describing the three elastic channels involved in the process ($a(b+2)+A \rightarrow f(=b+1)+F(=A+1) \rightarrow b + B(=A+2)$), one can stringently test two-nucleon transfer reaction theory.

It is found that a reaction description, which takes into account the successive and simultaneous contributions within the framework of the second order Distorted Wave Born Approximation (DWBA), properly corrected by non-orthogonality effects (G. Potel, private communication \cite{Potel:unp}, cf. also \cite{Potel:13b} concering the formalism), provides a quantitative descriptions of Cooper pair transfer, within uncertainties lying  below the $10\%$ level.

These results testify to the fact that pairing studies and the way they are revealed through two-nucleon transfer reactions have become quantitative, allowing to systematically  compare, within experimental  errors, theory with observations in terms of absolute differential cross sections and not only relative ones as done before.

\section{Cooper pairs}

The pairing interaction correlates pairs of nucleons moving in time reversal states over lengths of the order of $\xi=\hbar v_F/E_{corr}$, much larger than nuclear dimensions (cf. e.g. \cite{Bohr:75}), in keeping with the fact that the associated two--nucleon correlation energy is $E_{corr}\approx$ 0.5--2 MeV.
These extended, strongly overlapping virtual objects, known as Cooper pairs, affect most of the properties of nuclei close
to their ground state, as well as of their decay.
A textbook example of this last assertion is provided by the exotic decay $^{223}$Ra $\to ^{209}$Pb+ $^{14}C$. The measured decay constant 
$\lambda= 4.3 \times 10^{-16}$s${^{-1}}$,
implies that the wavefunction describing the ground state of the superfluid nucleus $^{223}$Ra has a component of amplitude of about $10^{-5}$
corresponding to a shape closely resembling $^{209}$Pb in contact with $^{14}$C. 
But this requirement can be fulfilled only if this exotic, strongly deformed system, is superfluid. In other words, if pairs of nucleons are correlated over 
distances of the order of 20 fm, sum of the Pb and C diameters \cite{Barranco:88}.

\subsection{Pairing correlations}

Nuclear superfluidity can be studied at profit in terms  of the mean field, BCS diagonalization
of the pairing Hamiltonian \cite{Bardeen:57both}, namely,
\begin{equation}
H = H_{sp} + V_p,
\label{H}
\end{equation}
where
\begin{equation}
H_{sp} = \sum_{\nu} (\epsilon_{\nu} - \lambda) a^+_{\nu} a_{\nu},
\label{Hsp}
\end{equation}
while 
\begin{equation}
V_p = - \Delta (P^+ + P) - \frac{\Delta^2}{G},
\label{Vp}
\end{equation}
and
\begin{equation}
\Delta = G \alpha_0,
\label{delta}
\end{equation}
is the pairing gap ($\Delta \approx$ 12 MeV/$\sqrt{A}$), $G$ ($\approx 25$ MeV/$A$ ) being the pairing coupling constant \cite{Bohr:75},
and 
\begin{equation}
P^+ = \sum_{\nu>0} P^+_{\nu}= \sum_{\nu>0} a^+_{\nu}a^+_{\bar \nu},
\label{P+}
\end{equation}
\begin{equation}
P = \sum_{\nu >0} a_{\bar \nu} a_{\nu},
\label{P-}
\end{equation}
are the pair addition and pair removal  operators, $a_{\nu}$ and $a^+_{\nu}$  being single-particle  creation  and annihilation  operators,
$(\nu \bar \nu)$ labeling pairs of time reversal states.

The BCS ground state wavefunction describing the most favorable configuration  of pairs to profit from the pairing interaction, can be 
written in terms  of the product of the occupancy probabilities $h_{\nu}$ for individual pairs,
\begin{equation}
|BCS> = \Pi_{\nu} ( (1 - h_{\nu})^{1/2} + h_{\nu}^{1/2} a^+_{\nu}a^+_{\bar \nu}) |0>,
\end{equation}
where $|0>$ is the fermion vacuum.

Superfluidity is tantamount to the existence of a finite average value of the operators  (\ref{P+}), (\ref{P-})
in this state, that is, to a finite value of the order parameter
\begin{equation}
\alpha_0 = <BCS|P^+|BCS> = <BCS|P|BCS>^*,
\end{equation}
 which is equivalent to Cooper pair condensation. In fact, $\alpha_0$ gives  a measure of the 
number of correlated pairs in the BCS ground state.
While the pairing gap (\ref{delta}) is an important quantity relating theory with experiment, $\alpha_0$ 
provides the specific measure  of superfluidity. In fact, the matrix elements of the pairing interaction
may vanish for specific regions of space,  or in the case of specific pairs of time reversal orbits, but this does not necessarily
imply a vanishing of the order parameter $\alpha_0$, nor the obliteration of superfluidity.

In keeping with the fact that Cooper pair tunneling is proportional to $|\alpha_0|^2$, this quantity plays also the
role of a $(L=0)$ two-nucleon
transfer sum rule, sum rule which is essentially exhausted by the superfluid nuclear $|BCS>$ ground state (see Fig. 3). Within the above context, one can posit that two-nucleon transfer reactions are the specific probes of pairing in nuclei.   

\subsection{Fluctuations}
The BCS solution of the pairing Hamiltonian was recasted by Bogoliubov \cite{Bogoliubov:58} and Valatin \cite{Valatin:58} in terms of quasiparticles, 
\begin{equation}
\alpha^+_{\nu} = U_{\nu} a^+_{\nu} - V_{\nu} a_{\bar \nu},
\end{equation}
linear transformation inducing the rotation in  $(a^+,a)$-space which diagonalizes  the Hamiltonian (\ref{H}).

The variational parameters $U_{\nu},V_{\nu}$ appearing in the above
relation indicate that $\alpha^+_{\nu}$ acting on $|0>$ creates a particle 
in the state $|\nu>$ which is empty with a probability $U^2_{\nu} \equiv (1 -h_{\nu})$, and annihilates a particle in the time reversal state $|\bar \nu>$
(creates a hole) which is occupied with probability $V_{\nu}^2 (\equiv h_{\nu})$. Thus, 
\begin{equation}
|BCS> = \Pi_{\nu>0} (U_{\nu} +V_{\nu} a^+_{\nu}a^+_{\bar \nu}) |0>,
\end{equation}
is the quasiparticle vacuum, as $|BCS> \sim \Pi_{\nu} \alpha_{\nu} |0>$, the order parameter being 
\begin{equation}
\alpha_0 = \sum_{\nu>0} U_{\nu}V_{\nu}.
\label{UV}
\end{equation}
Making use of these results we collect in Table 1  the spectroscopic amplitudes associated with  the reactions  
$^{A+2}$Sn(p,t)$^A$ Sn, for $A$ in the interval 112-126, as well as the spectroscopic amplitudes of other pairing modes (see below as well as \cite{Barranco:01,Gori:04}).

The BCS number and gap equations are,
\begin{equation}
N = 2 \sum_{\nu>0} V_{\nu}^2,
\label{numb}
\end{equation}  
\begin{equation}
\frac{1}{G} = \sum_{\nu>0} \frac{1}{2 E_{\nu}},
\end{equation}
where 
\begin{equation}
V_{\nu} = \frac{1}{\sqrt{2}} \left( 1 - \frac{\epsilon_{\nu} - \lambda}{\epsilon_{\nu}} \right)^{1/2},
\end{equation}
\begin{equation}
U_{\nu} = \frac{1}{\sqrt{2}} \left( 1 + \frac{\epsilon_{\nu} - \lambda}{\epsilon_{\nu}} \right)^{1/2},
\end{equation}
while the quasiparticle energy  is defined as 
\begin{equation}
E_{\nu} = \sqrt{(\epsilon_{\nu} - \lambda)^2 + \Delta^2}.
\label{eqp}
\end{equation}

\subsection{Pairing rotations}

The phase of the ground state BCS wavefunction may be chosen so that $U_{\nu} = |U_{\nu}| = U'_{\nu}$
is real and $V_{\nu} = V_{\nu}' e^{2 i \phi}$ $(V'_{\nu} \equiv |V_{\nu}|)$. Thus \cite{Bardeen:57both,Schrieffer:64,Schrieffer:72},
\begin{equation}
|BCS(\phi)>_{\cal K}  = \Pi_{\nu>0} (U'_{\nu} + V'_{\nu} e^{-2 i \phi} a^+_{\nu} a^+_{\bar \nu}) |0> = 
\Pi_{\nu>0} (U'_{\nu} + V'_{\nu} a^{'+}_{\nu} a^{'+}_{\bar \nu}) |0> = |BCS(\phi =0)>_{\cal K'},
\label{mean}
\end{equation}
where $a^{'+}_{\nu} = e^{-i \phi}a^+_{\nu}$ and   $a^{'+}_{\bar \nu} = e^{-i \phi}a^+_{\bar \nu}$.
This is in keeping with the fact that $a^+_{\nu}$ and $a^+_{\bar \nu}$ are single-particle  creation 
operators which under gauge transformations (rotations
in the 2D-gauge space of angle $\phi$) induced by the operator $G(\phi) = e^{- i \hat N (\phi)}$ and connecting the intrinsic and the laboratory frames of reference ${\cal K}$ and ${\cal K'}$ respectively, behave according to 
$a^{'+}_{\nu} = {\cal G} (\phi) a^+_{\nu} {\cal G}^{-1} (\phi)= e^{- i \phi} a^+_{\nu}$ and 
$a^{'+}_{\bar \nu} = {\cal G} (\phi) a^+_{\bar \nu} {\cal G}^{-1} (\phi)= e^{- i \phi} a^+_{\bar \nu}$, a consequence of the fact that $\hat N$ is the number operator and that $[\hat N, a^+_{\nu}] = a^+_{\nu}$.

The fact that the  mean field ground state  ($|BCS(\phi)\rangle_{\cal K}$) is a product of operators - one for each pair state - acting on the vacuum,
implies that (\ref{mean}) represents an ensemble of ground state wavefunctions averaged over systems with $... N-2,N,N+2 ...$ even number of particles.
In fact, (\ref{mean}) can also be written in the form 


\begin{equation}
|BCS>_{\cal K} = \left( \Pi_{\nu>0} U'_{\nu} \right ) 
( 1 + ... + 
\frac{e^{-(N-2)i \phi}}{\left(\frac{N-2}{2}\right)!} 
\left( \sum_{\nu>0} c_{\nu} a^+_{\nu}a^+_{\bar \nu} \right)^{\frac{N-2}{2}} +  \\
\frac{e^{-Ni \phi}}{\left(\frac{N}{2}\right)!} 
\left(\sum_{\nu>0}c_{\nu} a^+_{\nu}a^+_{\bar \nu}\right)^{\frac {N}{2}}   \\
\nonumber
\end{equation}
\begin{equation}
 + \frac{e^{-(N+2)i \phi}}{\left(\frac{N+2}{2}\right)!} 
\left(\sum_{\nu>0} c_{\nu} a^+_{\nu}a^+_{\bar \nu} \right)^{\frac {N+2}{2}} + ... 
)|0\rangle,
\end{equation} 
where $c_{\nu} = V'_{\nu}/U'_{\nu}$.

Adjusting the Lagrange multiplier $\lambda$ (chemical potential, see Eqs. (\ref{numb}-\ref{eqp})), one can ensure that the mean number of fermions 
(Eq. (\ref{numb})) has the desired value $N_0$.
Summing up, the BCS ground state is a wavepacket in the number of particles. In other words, it is a deformed state in gauge space  defining a privileged 
orientation in this space, and thus an intrinsic coordinate system ${\cal K'}$ \cite{Anderson:58, Bohr:88,Bes:66}.
The magnitude of this deformation is measured by $\alpha_0$.

\subsection{Pairing vibrations}
 
All the above arguments, point to a static picture of nuclear superfluidity which results from BCS theory. This is quite 
natural, as one is dealing with a mean field approximation.
The situation is radically changed  taking into account the interaction 
acting among the Cooper pairs (quasiparticles) which has been neglected until now, that is the term
$- G (P^+ -\alpha_0)(P-\alpha_0)$ left out in the mean field (BCS) approximation leading to (\ref{Vp}) \cite{Anderson:58,Bes:66}.
This interaction can essentially be written as (for details see e.g. \cite{Brink:05} App. J)
\begin{equation}
H_{residual} = H^{'}_p + H^{''}_p,
\end{equation} 
where 
\begin{equation}
H^{'}_p = - \frac{G}{4} 
\left( \sum_{\nu>0} (U^2_{\nu} - V^2_{\nu})(P^+_{\nu} + P_{\nu}) \right)^2,
\end{equation}
and 
\begin{equation}
H_p^{''} = \frac{G}{4} \left( \sum_{\nu>0} (P^+ - P) \right)^2.
\end{equation}
The term $H'_p$ gives rise to vibrations of the pairing gap  which (virtually) change particle number in $\pm$ 2 units. The energy
of these pairing vibrations cannot be lower than 2$\Delta$. They are, as a rule, little collective, corresponding  essentially 
to almost pure two-quasiparticle excitations (see excited $0^+$ states of Fig. 3).

The term $H_p^{''}$ leads to a solution of particular interest, displaying exactly zero energy, thus being degenerate with
the ground state. The associated wavefunction is proportional to the particle number operator and thus to the gauge operator inducing 
an infinitesimal rotation in gauge space. The fluctuations associated with this zero frequency mode diverge, although the Hamiltonian 
defines a finite inertia. 
A proper inclusion of these  fluctuations (of the orientation angle $\phi$ in gauge space) restores gauge invariance in the $|BCS(\phi)>_{\cal K}$
state leading to states with fixed particle number 
\begin{equation}
|N_0\rangle \sim \int_0^{2 \pi} d \phi e^{i N_0 \phi} |BCS(\phi)>_{\cal K} \sim 
(\sum_{\nu>0} c_{\nu} a^+_{\nu}a^+_{\bar \nu})^{N_0/2} |0>.
\end{equation}
These are the members of the pairing rotational band, e.g. the ground states of the superfluid Sn-isotope nuclei. These states provide 
the nuclear embodiment of Schrieffer's ensemble of ground state wavefunctions which is at the basis of the BCS theory of superconductivity. 

Summing up, while the correlations associated with $H_p^{''}$ lead to divergent fluctuations  which eventually restore symmetry
($[H_{sp} + H_p^{''}, \hat N] =0$),
$H_p'$ gives essentially rise  to non-collective particle number fluctuations, which are essentially pure two-quasiparticle states.

The situation is quite different in the case of normal systems, where pairing vibrations, namely the pair addition and pair removal modes \cite{Bohr:64,Bohr:75,Bes:66},
are strongly excited in two--particle  transfer reactions (see Fig. 4). Let us elaborate on this point, making use of the so called two-level 
model, in wihch the single-particle levels associated  with occupied (empty) states are assumed to be degenerate and separated by an energy $D$. 
The parameter which measures the interplay between pairing correlations and shell effects is
\begin{equation}
x = \frac{2 G \Omega}{D},
\end{equation}
where $G (\approx 25 $/A MeV) is the pairing coupling constant , while $\Omega = (2 j+1)/2$ is the pair degeneracy
of each of the two levels, assumed to be equal. Making use of the simple estimate of the spacing $D=2/\rho$ between levels close to the Fermi energy
in terms of the level density  (for one type of nucleons, e.g. neutrons) $\rho = 3A/2 \epsilon_F$ \cite{Bohr:69}, one obtains $D= 4 \epsilon_F/3A$ which 
 for $^{120}$Sn leads to $D \sim 0.4 MeV$ (cf. e.g. \cite{Brink:05} Ch. 2 and refs therein). Making use of the fact that the average pair degeneracy of the valence orbitals of $^{120}$Sn is
approximately 3,  one obtains $x \approx 3$, implying that pairing  effects overwhelm shell effects  and the static (pairing rotational) 
view of Cooper pair condensation is operative. 

On the other hand, in a closed shell system like, e.g., $^{208}$Pb, $D \approx 3.4$ MeV. Making use of the fact that the last occupied orbit
in $^{206}$Pb is a $p_{1/2}$ orbit, the first unoccupied being a $g_{9/2}$ level, one can use $\Omega =5$ in calculating  the value of $x$, which
becomes $x =0.35$. This value indicates that, in the present case,  shell effects are dominant. 

This does not mean that Cooper pairs are not present in the ground state  of $^{208}$Pb. It means that  they break as soon as they are created as virtual states through ground state correlations. Testifying to this scenario is the fact that  the expected $2p-2h$ $0^+$ state in $^{208}$Pb  at an energy of $ 2 D  \approx 6.8$ MeV, is observed at
4.9 MeV. The difference between these two numbers corresponds almost exactly  to the sum of the  correlation energy of $^{208}$Pb (gs) (0.640 MeV)
and of $^{210}$Pb (gs) (1.237 MeV). Thus, the first $0^+$ excited state of $^{208}$Pb corresponds to a two-phonons pairing vibrational state, product of the pair addition ($|^{210}$Pb$(gs)\rangle$) and of the pair removal ($|^{206}$Pb$(gs)\rangle$) modes of ($|^{208}$Pb$(gs)\rangle$).

In other words, we are in  presence of an incipient attempt of condensation in terms of two  correlated Cooper pairs, which, forced to be separated  
in two different nuclides
by particle conservation, get together in the highly correlated two-phonon pairing vibrational state of $^{208}$Pb. 
No surprise that  its microscopic structure can also be, rather easily, calculated almost exactly, by diagonalizing a schematic pairing force 
$H_p = - G P^+P$, in the harmonic approximation (RPA). The two-nucleon transfer spectroscopic amplitudes associated with the  reactions 
$^{206}$Pb(t,p)$^{208}$Pb(gs), $^{206}$Pb($^{18}$O,$^{16}$O)$^{208}$Pb(gs), $^{48}$Ca(t,p)$^{50}$Ca(gs), $^{10}$Be(p,t)$^8$Be(gs) and $^9$Li(p,t)$^7$Li(gs), and thus with the excitation (deexcitation) of pair addition and pair subtraction modes, are collected in Table 1. 
For details, cf. \cite{Broglia:73a}. 

\section{Nuclear Field Theory}

Elementary modes of nuclear excitation, namely rotations, vibrations and single-particle motion constitute a choice  basis for a compact and
economic description  of the nuclear structure, as these states incorporate many of the nuclear correlations \cite{Bohr:75}. The price 
to be paid for using a correlated basis, is that  these modes are non-orthogonal to each other, reflecting the fact that all the nuclear degrees of freedom  are already
exhausted  by the single-particle  degrees of freedom. In fact, nuclear elementary modes of excitation are not free, but 
interact with each other  through a coupling term ($H_c$) linear in the single--particle and collective coordinates.
Within this (non-orthogonality)  context we
refer to the similar situation encountered in the case of two-particle transfer reaction mechanism (see above as well as App. A).

A self-consistent  field theoretical treatment of this coupling can provide, to each  order of perturbation chosen, the solution of the nuclear many-body problem, eliminating in the process overcompleteness and Pauli principle violation of the basis states.
Propagating these interactions and corrections to infinite order through the Dyson (Nambu-Gor'kov \cite{Gorkov:59} in the superfluid case) equation, one can carry out
a full diagonalization of the many-body problem,  provided the rules of the game have been worked out and proved to be right.

A possible answer to such a quest was provided by the Nuclear Field Theory (NFT) \cite{Bes:76a,Bes:76b,Bes:76c,Bes:75,Mottelson:76,Bortignon:77,Broglia:76}. With four rules which allow to select the interactions (particle--vibration coupling and four--point vertices $v$), calculate the collective  (e.g. RPA for vibrations in spherical, normal nuclei) and the single--particle degrees of freedom (mean field associated with $v$), and  eliminate  non--allowed processes (diagrams),
it was possible to prove  that  NFT propagators provided the same answer as the Feynman-Goldstone (F-G) many--body propagators to any order of perturbation (cf. App. B). Such a proof is tantamount to the statement that the use of NFT techniques for solving   nuclear many-body problems, lead to the correct solution to any given order of perturbation. In particular to the exact solution in the case in which the different contributions are summed to infinite order.

Within this  scenario, and provided one has experimental information concerning the collective  modes which  dress the single-particle 
states and  renormalize the nucleon-nucleon  (pairing) interaction,  NFT is expected to provide  accurate predictions concerning the nuclear structure
in general, and two-nucleon  spectroscopic amplitudes in particular.

This is in fact the case concerning $^{11}$Li and $^{12}$Be, two systems studied in much detail with the help of NFT. The two-nucleon transfer 
spectroscopic amplitudes associated with the reactions $^1$H($^{11}$Li,$^9$Li($J^{\pi}))^3$H ($J^{\pi} = 3/2^- $ (gs) and $J^{\pi} = 1/2^-$ ($E^*=$ 2.69 MeV)),
$^7$Li(t,p)$^9$Li(gs) and $^{10}$Be(t,p)$^{12}$Be(gs) are collected in Table 1.
For details of the calculations of the  wavefunctions we refer to \cite{Barranco:01,Gori:04}.

\section{Reaction mechanism}

Through the single action of the nucleon-nucleon interaction on one nucleon, a two-particle transfer reaction can take place, the second particle being able to  follow the first one  profitting of the particle-particle (Cooper pair) correlation. Although this is correct, it does not reflect the whole story. To explain why this is so, let us make use of the independent particle model,
in which case nucleons do not interact with each other but because of Pauli principle. Nonetheless, also in this limit 
two-particle transfer can still take place as a first order process in the nucleon-nucleon interaction,  the second nucleon going over, profitting this time  of the fact that the  single-particle wavefunctions are not orthogonal. But this result depends on the single-particle basis used and can thus hardly be correct.

 In fact, within the framework of the two-center shell model basis of reaction theory (cf. e.g. \cite{Maruhn:72}), the non-orthogonality term is by definition zero. This is in keeping  with the fact that the overlaps $\langle \phi^{(b)}_{a'_1}|\phi^{(A)}_{a_1} \rangle = \langle \phi^{(b)}_{a'_2}|\phi^{(A)}_{a_2} \rangle = 0$. Thus, in the independent particle limit,  two-particle simultaneous transfer, in which $v$ acts only once, is non operative. 

Consequently, in a non-orthogonal basis like the one used in the present paper, the above parlance implies that, in lowest order 
in the nucleon-nucleon interaction, there must be two contributions: the simultaneous transfer and another one known as non-orthogonality
correction, which  exactly cancels  the first order in the independent particle limit (cf. e.g. \cite{Broglia:04a} p.424).
Because in nuclei, the correlations between Cooper pair partners are weak, cancellation is rather strong, and successive transfer, is, as a rule, 
the main contribution to two-particle transfer in lowest (second) order in the nucleon-nucleon interaction.
It constitutes a further check of the consistency of such a description that in the limit of strong correlations between Cooper pair partners, the non-orthogonality correction is needed  to ensure that successive transfer vanishes as the Cooper binding energy becomes large, ensuring simultaneous transfer to be the dominant, lowest order transfer mechanism of the bosonic dinuclear system (see App. A). 

The theory of two-nucleon transfer reactions involving superfluid nuclei essentially started with the work of  Yoshida 
\cite{Yoshida:62} who pointed out  that the differential cross section connecting the ground states of two superfluid nuclei is proportional to 
$(\Delta/G)^2$, that is $d \sigma_{gs} \to gs \sim (\alpha_0)^2$. Shortly after, Glendenning \cite{Glendenning:65} provided a microscopic description  in terms of effective formfactors and DWBA, followed by the seminal work of Bayman \cite{Bayman:68}. In all these cases  use was made of the zero-range  approximation (no-recoil effects) and simultaneous transfer was the only process considered.

Work started in the seventies and carrying on to our days (cf. 
\cite{Udagawa:73,Potel:13b,Chien:75,Segawa:75,Schneider:76,Gotz:75,Takacsy:74,Hashimoto:78,Kubo:78,Yagi:79,Igarashi:91,Becha:97,Tanihata:08,Potel:10,Bayman:82,Glendenning:unp,Broglia:Lectures,Glendenning:63,Bayman:67,Broglia:67,Satchler:64,Satchler:83,Austern:70,Flynn:70b,Ball:71,Hering:70,Bayman:70,Bayman:71,Riedel:68,Barz:69,Franey:78,Maglione:85,Krouglov:00,Krouglov:01,Alkhazov:02,Keeley:07a,Potel:11PRL,Glendenning:69,Ascuitto:69,Ascuitto:70,Tamura:70,Ascuitto:71,Ascuitto:72,Thompson:88,Potel:13a,Schaeffer:72,Toyama:72,Kunz:73,Broglia:73c,Maglione:87,Timofeyuk:00,Wimmer:10,Mougeot:12,Thompson:13,Tanihata:13,Bildstein:12,Fortune:12} and App. C),
has provided the basis, developed the formalism and worked out the
tests for a consistent 2nd order DWBA description  of two-nucleon transfer process.

Making use of an implementation (\cite{Potel:unp}, cf. also ref. \cite{Potel:13b,Bayman:82}) of such second order DWBA which includes successive and simultaneous transfer, corrected from non-orthogonality, of the spectroscopic amplitudes collected in Table 1,
and of global optical parameters from the literature (see Table 2), a number of two-nucleon transfer absolute differential cross sections have been calculated. They are compared with the experimental findings in Figs. 5 and 6. In Table 3
the corresponding integrated absolute cross sections are collected in comparison with observations. It is fair to state that theory provides an overall account of the experimental findings.

Let us conclude this paper by mentioning that throughout it we have dealt, essentially on equal footing, with light (i.e. $(p,t)$ and $(t,p)$) and heavy ion (e.g. $(^{18}$O,$^{16}$O$)$) reactions, without much further ado.
This is in keeping with the fact that the formalism to deal with each of this type of reactions, as well as to work out the connection and differences between them, is well known in the literature (cf. e.g. \cite{Broglia:73a,Potel:13b,Broglia:04a,Bayman:82,Becha:97,Thompson:88}, and refs. therein).

\section{Conclusions}

With the help of spectroscopic amplitudes which provide an accurate description of the structure, in particular the pairing 
aspects of it, of nuclei covering essentially the whole mass table,  and of empirical optical potentials, we have 
tested two-nucleon transfer reaction theory. Second order DWBA which includes simultaneous  and successive transfer properly corrected with non-orthogonality effects, provides a quantitative account of the absolute differential cross sections,
essentially within experimental errors and below the 10\% level.

\section{Acknowledgment} 

Discussions and collaboration with B. Bayman are gratefully acknowledged. We want also to thank L. Zetta and P. Guazzoni
for clarification and access regarding their systematic, state of the art (p,t) data.

\appendix

\section{Relative importance of successive and simultaneous transfer and non-orthogonality corrections}

In what follows we discuss the relative importance of successive and simultaneous two-neutron transfer and of non-orthogonality 
corrections associated with the reaction 

\begin{equation}
\alpha \equiv  a(=b+2) + A \to b + B(=A+2) \equiv \beta ,
\label{A1}
\end{equation}
in the limits of independent particles and of strongly correlated Cooper pairs, making use, for simplicity, of the semiclassical approximation (for details cf. \cite{Broglia:04a},\cite{Broglia:Lectures}  and refs. therein). In this case the two-particle transfer differential cross section can be written as,
\begin{equation}
\frac{d \sigma_{\alpha \to \beta} }{d \Omega} = P_{\alpha \to \beta} (t = +\infty) 
\sqrt{ \left( \frac{d \sigma_{\alpha}}{d \Omega} \right)_{el} }
\sqrt{ \left( \frac{d \sigma_{\beta}}{d \Omega} \right)_{el}}\qquad, 
\label{A2}
\end{equation}
where $P$ is the absolute value squared of a quantum mechanical transition amplitude. It gives the probability that the system at $t = + \infty$ is found in the final channel. The quantities $(d \sigma/d\Omega)_{el}$ are the classical elastic cross sections  in the center of mass system, calculated in terms of the deflection function, namely the functional relating the impact parameter and the scattering angle. 

The transfer amplitude can be written as

\begin{equation}
a(t = + \infty) = a^{(1)}(\infty) - a^{(NO)}(\infty) + \tilde a^{(2)} ( \infty),
\label{A3}
\end{equation}
where $\tilde a^{(2)}(\infty)$ labels  the successive transfer amplitude expressed in the post-post representation (see below).
The simultaneous transfer amplitude is given by (see Fig. 7(I))

\begin{eqnarray}
a^{(1)} (\infty) = \frac{1}{i \hbar} \int^{\infty}_{-\infty} dt (\psi^b \psi^B, (V_{bB} - <V_{bB}>) \psi^a \psi^A ) \times 
{\rm exp} [\frac{i}{\hbar} (E^{bB} - E^{aA}) t] \nonumber \\
\approx \frac{2}{i \hbar} \int^{\infty}_{- \infty}  dt \left( \phi^{B(A)} (S^B{(2n)}; \vec r_{1A}, \vec r_{2A}), U(r_{1b}) 
e^{i (\sigma_1 + \sigma_2)}
\phi^{a(b)} (S^a{(2n)}; \vec r_{1b}, \vec r_{2b}) \right) {\rm exp} [\frac{i}{\hbar} (E^{bB} - E^{aA}) t + \gamma(t)],
\end{eqnarray}
where 
\begin{equation}
\sigma_1 + \sigma_2 = \frac{1}{\hbar} \frac{m_n}{m_A} ( m_{aA} \vec v_{aA} (t) + m_{bB} v_{bB}(t)) \cdot (\vec r_{1\alpha}
+ \vec r_{2 \alpha}),
\end{equation}
in keeping with the fact that ${\rm exp} ( i (\sigma_1 + \sigma_2))$ takes care of recoil 
effects (Galilean transformation associated with the mismatch between entrance and exit channels), the phase $\gamma (t)$ being related  to the effective $Q-$value of the reaction. 

In the above expression, $\phi$ indicates an antisymmetrized, correlated two-particle (Cooper pair)  wavefunction, $S(2n)$ being the two-neutron separation energy (see Fig. 8), $U(r_{1b})$ being the single particle potential generated by nucleus $b$ ($U(r) = \int d^3 r' \rho^b(r') v(|r-r'|)$). The contribution arising from non-orthogonality effects can be written as (see Fig. 7(II))

\begin{align}
a^{(NO)} (\infty) = & \frac{1}{i \hbar} \int^{\infty}_{-\infty} dt (\psi^b \psi^B, (V_{bB} - <V_{bB}>) \psi^f \psi^F )
(\psi^f\psi^F, \psi^a \psi^A) 
{\rm exp} [\frac{i}{\hbar} (E^{bB} - E^{aA}) t]  \nonumber \\
&\approx \frac{2}{i \hbar} \int^{\infty}_{- \infty} \textrm{d}t (\phi^{B(F)} (S^B{(n)}; \vec r_{1A}), U(r_{1b}) 
e^{i \sigma_1}
\phi^{f(b)}(S^f(n); \vec r_{1b}))  \phi^{F(A)}  \nonumber\\
&\times (\phi^{F(A)}(S^F(n);\vec r_{2A}) e^{i \sigma_2}\phi^{a(f)}(S^a(n);\vec r_{2b})) 
{\rm exp} [\frac{i}{\hbar} (E^{bB} - E^{aA}) t + \gamma(t)], \notag \\ 
\end{align}
the reaction channel $f= (b+1) + F(=A+1)$ having been introduced, the quantity $S(n)$ being the one-neutron separation 
energy (see Fig. 7). The summation over $f(\equiv a'_1,a'_2)$ and $F (\equiv a_1,a_2)$ involves a restricted number of states, namely the valence shells in nuclei $B$ and $a$ respectively.

The successive transfer amplitude  $\tilde a^{(2)}_{\infty}$ written making use of the post-post representation is equal to 
(see Fig. 7(III))

\begin{eqnarray}
\tilde a^{(2)} (\infty) = \frac{1}{i \hbar} \int^{\infty}_{-\infty} dt (\psi^b \psi^B, (V_{bB} - <V_{bB}>) e^{i \sigma_1} \psi^f \psi^F ) \times 
{\rm exp} [\frac{i}{\hbar} (E^{bB} - E^{fF}) t + \gamma_1(t)] \nonumber  \\
\times \frac{1}{i \hbar} \int^{t}_{-\infty} dt' (\psi^f \psi^F, (V_{fF} - <V_{fF}>) e^{i \sigma_2} \psi^a \psi^A > \times 
{\rm exp} [\frac{i}{\hbar} (E^{fF} - E^{aA}) t' + \gamma_2(t)].
\end{eqnarray}

To gain insight into the  relative importance of the three terms contributing to Eq. (\ref{A3}) we discuss two situations, namely,
the independent-particle and the strong-correlation limits.

\subsection{Independent particle limit}

In the independent particle limit, the two transferred particles do not interact among themselves but for antisymmetrization. 
Thus, the separation energies fulfill the relations (see Fig. 8)
\begin{equation}
S^B(2n) = 2 S^B(n) = 2S^F(n),
\end{equation}
and 
\begin{equation}
S^a(2n) = 2 S^a(n) = 2 S^f(n).
\end{equation}
In this case 
\begin{equation}
\phi^{B(A)} (S^B(2n); \vec r_{1A},\vec r_{2A}) = \sum_{a_1 a_2} \phi_{a_1}^{B(F)} (S^B(n);\vec r_{1A}) 
\phi_{a_{2}}^{F(A)} (S^F(n);\vec r_{2a}),
\end{equation}
and 
\begin{equation}
\phi^{a(b)} (S^a(2n); \vec r_{1b},\vec r_{2b}) = 
\sum_{a^{'}_{1} a^{'}_{2}} \phi_{a^{'}_1}^{a(f)} (S^a(n);\vec r_{2b}) 
\phi_{a^{'}_{2}}^{f(b)} (S^f(n);\vec r_{1b}),
\end{equation}
where $(a_1, a_2) \equiv F$ and $(a'_1, a'_2) \equiv f$ span, as mentioned above, shells in nuclei $B$ and $a$ respectively. 

Inserting (A9) and (A10) in (A4) one can show that 
\begin{equation}
a^{(1)} (\infty) = a^{(NO)}(\infty).
\end{equation}
It can be shown  that, within the present approximation, $Im \; \tilde a^{(2)} =0$ and  
\begin{eqnarray}
\tilde a^{(2)} (\infty) = \frac{1}{i \hbar} \int^{\infty}_{-\infty} dt (\psi^b \psi^B, (V_{bB} - <V_{bB}>) e^{i \sigma_1} \psi^f \psi^F > \times 
{\rm exp} [\frac{i}{\hbar} (E^{bB} - E^{fF}) t + \gamma_1(t)] \nonumber  \\
\times \frac{1}{i \hbar} \int^{\infty}_{-\infty} dt' (\psi^f \psi^F, (V_{fF} - <V_{fF}>) e^{i \sigma_2} \psi^a \psi^A ) \times 
{\rm exp} [\frac{i}{\hbar} (E^{fF} - E^{aA}) t' + \gamma_2(t')].
\label{A12}
\end{eqnarray}
The total absolute differential cross section (\ref{A2}), where $P = |a(\infty)|^2 = |\tilde a^{(2)}|^2$, is then equal to the product of two one-particle transfer cross sections (see Fig. 9), associated with the (virtual) reaction channels
\begin{equation}
\alpha \equiv a+A \to f +F \equiv \gamma,
\end{equation}
and 
\begin{equation}
\gamma \equiv f +F \to b+B \equiv \beta.
\end{equation}

In fact, Eq.(\ref{A12}) involves no time ordering and consequently the two processes above are completely independent of each other. 
This result was expected because being $v_{12}= 0$, the transfer of one nucleon cannot influence, aside form selecting the
initial state for the second step, the behaviour of the other nucleon.

\subsection{Strong correlation (cluster) limit}

The second limit to be considered is the one in which the correlation betwen the two nucleons is so strong that (see Fig. 8)
\begin{equation}
S^a(2n) \approx S^a(n) >> S^f(n),
\label{A15}
\end{equation}
and 
\begin{equation}
S^B(2n) \approx S^B(n) >> S^F(n).
\label{A16}
\end{equation}
That is, the magnitude of the one-nucleon separation energy is strongly modified by pair breaking.

There is a different, although equivalent way to express (\ref{A3}) which is more convenient to discuss the strong coupling limit.
In fact, making use of the post-prior representation one can write
\begin{eqnarray}
a^{(2)}(t) = \tilde a^{(2)}(t) - a^{(NO)}(t) = 
\frac{1}{i \hbar} \int^{\infty}_{-\infty} dt (\psi^b \psi^B, (V_{bB} - <V_{bB}>) \psi^f \psi^F ) \nonumber \\ 
\times {\rm exp} [\frac{i}{\hbar} (E^{bB} - E^{fF}) t + \gamma_1(t)]\nonumber  \\
\times  \frac{1}{i \hbar} \int^{t}_{-\infty} dt' (\psi^f \psi^F, (V_{aA} - <V_{aA}>) \psi^a \psi^A ) 
{\rm exp} [\frac{i}{\hbar} (E^{fF} - E^{aA}) t' + \gamma_2(t')].
\end{eqnarray}
The relations  (\ref{A15}), (\ref{A16}) imply 

\begin{equation}
E^{fF} - E^{aA} = S^a(n) - S^F(n) >> \frac{\hbar}{\tau},
\end{equation}
where $\tau$ is the collision time. Consequently the real part of $a^{(2)}(\infty)$ vanishes exponentially  with the $Q-$value of the intermediate transition, while the imaginary part  vanishes inversely proportional to this energy.
One can thus write,
\begin{equation}
Re \;  a^{(2)} (\infty) \approx 0,
\end{equation} 
and 
\begin{equation}
a^{(2)}(\infty) \approx \frac{1}{i \hbar} \frac{\tau}{<E^{fF}> - E^{bB}} 
\sum_{fF} (\psi^b \psi^B, 
(V_{bB}- <V_{bB}>) 
\psi^f\psi^F)_{t=0} \times 
(\psi^f\psi^F,(V_{aA} - <V_{aA}) \psi^a \psi^A)_{t=0},
\end{equation} 
where one has made use of the fact that $E^{bB} \approx E^{aA}$. For $v_{12} \to \infty$, $(<E^{fF}> - E^{bB}) \to \infty$
and, consequently, 

\begin{equation}
\lim_{v_{12} \to \infty} a^{(2)} (\infty) = 0.
\end{equation} 

Thus the complete two-nucleon transfer amplitude is equal, in the strong coupling limit, to the amplitude $a^{(1)} (\infty)$.

Summing up, only when successive transfer and non-orthogonal corrections are included in the description of the two-nucleon 
transfer process, does one obtain a consistent description of the process, which correctly converges to both the weak and 
the strong correlation limiting values.

\section{Equivalence between NFT and F-G propagators}
In what follows we briefly recall the proof of equivalence of NFT and of F-G propagators. In 
the case of the Tamm-Dancoff propagator of  particle-hole collective modes, calculated in terms of a separable interaction, 
such a proof  consisted in 
demonstrating (for details cf. \cite{Bes:76c,Bes:75,Bortignon:77}) that
\begin{equation}
(NFT) \;\;  \sum_n \frac{\Lambda_n^2}{\omega - W_n +i \eta} = 
\frac
{V^2 \sum_{ki} \frac{q^2(k,i)}{\omega - \epsilon_{ki} +i \eta}}
{1 + V \sum_{ki} \frac{q^2(k,i)}{\omega - \epsilon_{ki} +i \eta}}, \;\; (F-G)\label{equality}
\end{equation}  
where $\Lambda_n$ is the particle-vibration coupling strength and $W_n$ is the energy of the collective modes, $V$ is the four-point vertex, $\epsilon_{ki}$ 
the particle-hole  excitations and $q(k,i)$ the particle-hole matrix element of the operator $Q$ of the separable  interaction.

Because of the unitary transformation  relating the particle-hole  $(k,i)$ and collective degrees of freedom $(n)$ basis \cite{Bohr:75}, the dispersion relations
\begin{equation}
\frac{1}{V} = \sum_{ki} \frac{q^2(ki)}{\epsilon_{ki} - W_n},
\end{equation} 
\begin{equation}
V = \sum_n \frac{\Lambda_n^2}{\epsilon_{ki} - W_n},
\end{equation}
are operative,  thus implying the equality (\ref{equality}).

\section{Development of two-nucleon transfer as a quantitative spectroscopic probe}

The developments which are at the basis of the quantitative description of two-nucleon transfer reactions were carried out in the sixties \cite{Glendenning:63,Bayman:68,Glendenning:65,Bayman:67,Broglia:67} within the context of one-step (simultaneous transfer) DWBA (\cite{Satchler:64}, see also \cite{Satchler:83,Austern:70}), in particular in connection with $(t,p)$ and $(p,t)$ reactions.
The matrix element $\langle bB(=A+2)|V_f|a(=b+2)A\rangle$ of the effective interaction between the two neutrons and the proton contains all the nuclear structure information needed to describe the reaction process. It is a function of the center of mass coordinate $\vec R$ of the dineutron, and of the relative distance $\vec \rho$ between the center of mass of this system and of the incoming or outgoing proton, depending whether one is studying $(p,t)$ or $(t,p)$ reactions. This function is thus intimately related to the interweaving of finite range and recoil effects with nuclear structure correlations. The explicit evaluation of $\langle bB | V_f | aA \rangle$ leads to the effective two-nucleon (simultaneous) transfer form factor.

The method proposed by Glendenning \cite{Glendenning:65} employs single-particle harmonic oscillator wavefunctions and the Moshinsky transformation brackets, providing a convenient framework to study the coherence properties of the two-nucleon transfer reactions form factors.
Because for strongly absorbed particles, the two nucleon transfer cross section is approximately proportional to the square of the formfactor amplitude near the nuclear surface, the method of ref. \cite{Glendenning:65} was instrumental to provide a microscopic validation of the assertion that two-nucleon transfer is the specific probe of nuclear pairing.
A projection method to calculate the six-dimensional integral appearing in $\langle bB| V_f |aA \rangle$, more convenient to treat finite range and recoil effects, was introduced by Bayman and Kallio \cite{Bayman:67}.

These approaches allowed for a systematic and detailed comparison between experimental and theoretical relative cross sections. They were instrumental in helping to understand the coherence properties which are at the basis of nuclear correlations. Most outstanding pairing correlations \cite{Broglia:67}, but also particle-hole correlations (cf. e.g. \cite{Broglia:73a} and refs. therein).

Concerning the absolute value of the cross sections it was shown in \cite{Flynn:70b} that the experimental 
findings could be accounted for, within the framework of the zero-range approximation ($V_f(\vec \rho) \Phi_{(0S)}(\vec \rho) = D_0 \delta(\vec \rho)$, $D^2_0 = N'_0 W$, $W=10^4$ MeV$^2$fm$^3$, $N'_0 = (\sum \textrm{d}\sigma/\textrm{d}\Omega)_{exp}/(\sum \textrm{d}\sigma/\textrm{d}\Omega)_{th}$), using a common value $N'_0 = 23$, quite close to that reported in ref. \cite{Ball:71} in connection with the analysis of $(p,t)$ data on Zr isotopes. Using $(p,t)$ data on Ca, Sn, and Pb and a different set of optical parameters than those used in \cite{Flynn:70b} and \cite{Ball:71}, a value $N'_0 = 32$ was quoted in \cite{Broglia:72b}.

Theoretical estimates \cite{Hering:70,Broglia:72b} gave values of $N'_0$ near 1, i.e., about 25-30 times too small as compared to the empirical normalizations. The main reasons for this significant discrepancy were: a) the use of oversimplified triton wavefunctions, b) the use of the zero-range approximation, c) the use of only simultaneous transfer.
Shortcomings a) and b) were eliminated in the work of Bayman \cite{Bayman:70,Bayman:71}, who calculated the $L=0$ transition in the Ca nuclei, by numerical evaluation of the six-dimensional DWBA integral. He obtained for the case of $^{48}$Ca$(t,p)^{50}$Ca$(gs)$ a value of $N'_0=3.2$. That is, the theoretical cross section underpredicted the absolute value of the experimental cross section by a factor of three, in spite of the fact that a correlated wavefunction in which two neutron outside the $^{48}$Ca-core are allowed to correlate in the six valence orbital through a pairing force whose coupling constant was adjusted to reproduce the ground state correlation energy.

As a result of the work of a number of groups, in particular that of Bayman, starting in the seventies and carrying to our days 
\cite{Broglia:72b,Broglia:04a,Glendenning:65,Bayman:68,Udagawa:73,Chien:75,Segawa:75,Schneider:76,Gotz:75,Takacsy:74,Hashimoto:78,Kubo:78,Yagi:79,Igarashi:91,Becha:97,Tanihata:08,Potel:10,Bayman:82,Potel:unp,Glendenning:unp,Broglia:Lectures,Glendenning:63,Bayman:67,Broglia:67,Satchler:64,Satchler:83,Austern:70,Flynn:70b,Ball:71,Hering:70,Bayman:70,Bayman:71,Riedel:68,Barz:69,Franey:78,Maglione:85,Krouglov:00,Krouglov:01,Alkhazov:02,Keeley:07a,Potel:11PRL,Glendenning:69,Ascuitto:69,Ascuitto:70,Tamura:70,Ascuitto:71,Ascuitto:72,Thompson:88,Potel:13a,Schaeffer:72,Toyama:72,Kunz:73,Broglia:73c,Maglione:87,Timofeyuk:00,Wimmer:10,Mougeot:12,Thompson:13,Tanihata:13,Bildstein:12,Fortune:12}
we now know that the above discrepancy is mainly associated with the neglect of successive transfer and of non-orthogonality corrections (see Fig. 7).
In fact, making use of the same two-particle wavefunction to describe $|^{50}$Ca$(gs) \rangle$ used in ref. \cite{Bayman:82}, one obtains for the reaction $^{48}$Ca$(t,p)^{50}$Ca$(gs)$ an absolute differential cross section which reproduces the data within experimental errors (see Fig. 6  and Table 3).

In the early seventies, first attempts were made to take into account multi--step processes  in several contexts. Schaeffer and Bertsch \cite{Schaeffer:72} considered the $^4$He channel in the charge exchange reaction $^1$H$\longrightarrow ^4$He$\longrightarrow ^3$H. Igarashi et al. (1972) (cf. ref. \cite{Igarashi:91} and refs. therein) took into account the role of virtual triton formation in the one--neutron transfer reaction $^1$H$\longrightarrow ^3$H$\longrightarrow ^2$H. 
Successive transfer processes were considered by Toyama \cite{Toyama:72} , and the importance  of non-orthogonality terms was discussed in \cite{Udagawa:73, Kunz:73} for light-ion reactions,
based on the  second--order approximation to the Coupled Reaction Channels (CRC) equations for two--particle transfer.
A reaction formalism which included simultaneous, successive, and non--orthogonality terms within the context of two--nucleon transfer between heavy ions was presented in \cite{Broglia:73c,Gotz:75},
developing  a semiclassical approximation that has been successfully  applied  to a number of reactions
\cite{Franey:78, Maglione:85, Maglione:87}. 
The fully quantum treatment based on second-order DWBA  has been applied to the $^{208}$Pb($^{16}$O,$^{18}$O)$^{206}$Pb heavy ion reaction in  \cite{Bayman:82}. 
By describing the two valence neutrons of $^{18}$O with  a mixing of $d_{5/2}$ and $s_{1/2}$ configurations, theory  underestimated  experimental results  by a factor of two. 
These results may be compared with the calculations, reported in Table 3 and displayed in Fig. 5, where one has included a significant amount of $d_{3/2}$ mixing, thus improving the agreement with experimental data (see also \cite{Maglione:85}). In a very detailed paper, Igarashi et al. \cite{Igarashi:91} addressed the issue of  the relative importance of simultaneous and successive contributions to the two--step DWBA transfer amplitude in (p,t) reactions, developing the computer code TWOFNR and emphasizing the role of successive transfer in the reaction mechanism. Using a spin--dependent proton--neutron interaction both to generate the triton and deuteron wavefunction and the transfer potential, they were able to account for unnatural as well as for natural parity transitions absolute cross sections. They also studied the effect of the coupling to unbound states of the deuteron with the Discretized Continuum Channel method, concluding that their contribution to the total cross section is small.

More recently, two--nucleon transfer reactions have been analyzed within the CRC formalism, mostly with the computer code FRESCO (
\cite{Becha:97, Timofeyuk:00, Krouglov:00, Krouglov:01, Keeley:07a, Tanihata:08, Wimmer:10,Mougeot:12, Thompson:13}). 
In \cite{Timofeyuk:00}, the authors describe $p$--$^6$He scattering at very low energy (0.97 MeV in the CM frame) including explicitly elastic, inelastic and transfer channels. The interplay of the different channels appear to be essential for a correct description of the two--neutron transfer process.
Similar conclusions are drawn from the $^6$He--$^{12}$C scattering with a 5.9 MeV He beam \cite{Krouglov:00}, 
On the other hand, the analysis of the interference of elastic $^6$He--$^4$He scattering with the two--neutron transfer process at 151 MeV 
\cite{Krouglov:01} showed a much smaller influence of the excited $2^+$ state of $^6$He. 
The recent analysis of the $^{1}$H($^{11}$Li,$^{9}$Li)$^{3}$H reaction \cite{Tanihata:08,Tanihata:13,Potel:10} showed that data could be reproduced 
by a 2--step DWBA calculation with a  description of the structure of $^{11}$Li based on  an NFT wavefunction for $^{11}$Li 
\cite{Barranco:01}, with a proper inclusion of the continuum states in the intermediate $^{10}$Li--$d$ channel.
The calculation  reproduced the experimental yields of both the $3/2^-$ ground state and the first $1/2^-$ excited state of $^9$Li \cite{Potel:10}. In this work, the role of inelastic and breakup channels was estimated to be negligible. 
In \cite{Keeley:07a} and \cite{Mougeot:12}, the study of a complete set of data (elastic, one--neutron transfer and two--neutron transfer to ground and excited states) 
at 15.7 $A$ MeV for the system $^8$He--p was reported.  A  CRC analysis including the excited $2^+$ state of $^6$He and the breakup states of the deuteron provide an overall account of the experimental data. 
The availability of tritium targets, recently developed at ISOLDE (T-REX setup \cite{Bildstein:12}), will make  possible the study of  new (t,p) reactions. As an example, one can mention the absolute cross sections of the t($^{30}$Mg,$^{32}$Mg)p reaction, associated with the population of the ground state and for the first excited $0^+$ 
state \cite{Wimmer:10}. The interpretation of this  experiment may provide important information, concerning the wavefunctions of the two states, the role of intruder configurations 
and the  phenomenon of shape coexistence in this nucleus which belong to the so called "island of inversion" \cite{Fortune:12}. To be noticed that work on two--nucleon transfer reactions at sub--barrier Coulomb energies, as well as those which studied the influence of different $p-d$, $d-t$ channels for  particular light ions, e.g. $^8$He and $^6$He, and the excitation of unnatural parity states of heavy nuclei, have played a relevant role in the development of two--nucleon transfer processes as a quantitative probe \cite{Corradi:11,DeYoung:05,Chatterjee:08,Lemasson:10,Lemasson:11,Keeley:07,Charlton:76,Yasue:77,Takemasa:79,Kurokawa:87}.  

Recapitulating, two main inputs helped at making two-nucleon transfer reactions a quantitative tool to study nuclear correlations.
The first, was the fact that non-orthogonality corrections entered reaction theory, already at the level of first order DWBA (see e.g. \cite{Udagawa:73,Gotz:75} and refs. therein), in a similar way in which non-orthogonality of elementary modes of nuclear excitation leads to a first-order particle-vibration coupling vertex (\cite{Bohr:75}).
The second was associated with the study of multi-step processes in connection with direct nuclear reactions (Fig. 10). 

Very early in the study of two-nucleon transfer reactions in deformed nuclei \cite{Glendenning:69,Ascuitto:69,Ascuitto:70,Tamura:70,Ascuitto:71,Ascuitto:72}, let alone of one-particle transfer and inelastic processes, the need to consider second-order (multi-step) inelastic channels was clear.
Softwares developed in this connection were eventually modified to take successive transfer into account. Within this context one can mention the software FRESCO \cite{Thompson:88}, and e.g. its application to the detailed study of the $^{40}$Ca$(t,p)^{42}$Ca absolute reaction cross section \cite{Becha:97}, taking into account both one-step and two-steps reaction channels in a consistent manner. A recent and similar analysis of the inverse kinematic reaction $^{1}$H($^{11}$Li,$^{9}$Li)$^{3}$H \cite{Tanihata:08} providing important insight into the origin of pairing correlation in nuclei (cf. \cite{Potel:10} and \cite{Potel:13a} and refs. therein).
A number of such type of analysis, namely coupled reaction channel (CRC) method implemented in FRESCO, in particular concerning the properties of light, neutron reach nuclei have been reported in the literature during the last years, e.g. \cite{Krouglov:00,Krouglov:01,Keeley:07a}.

\newpage

%
%

\begin{table}[h!]
{\begin{tabular}{|c|c|c|c|c|c|c|c|}
\cline{1-8} 
a)& $^{112}$Sn & $^{114}$Sn&  $^{116}$Sn & $^{118}$Sn&  $^{120}$Sn &  $^{122}$Sn &  $^{124}$Sn          \\
\hline
1$d_{5/2}$            & 0.664      &  0.594   & 0.393    & 0.471      & 0.439     &  0.394    &  0.352                  \\
\hline 
0$g_{7/2}$            &  0.958     &  0.852  &  0.542     &  0.255   &  0.591      &  0.504  &   0.439                 \\
\hline 
2$s_{1/2}$            &  0.446    & 0.477    &  0.442    &  0.487     &  0.451   &  0.413     & 0.364                   \\
\hline 
1$d_{3/2}$            &  0.542    & 0.590   &  0.695    &  0.706     &  0.696   & 0.651   &   0.582                 \\
\hline 
0$h_{11/2}$            & 0.686     & 0.720    &  1.062     &  0.969     &  1.095   &  1.175    &   1.222                 \\
\hline 
\end{tabular}}
\end{table}

\begin{table}[h!]
{\begin{tabular}{|c|c|c|}
\cline{1-3} 
 b)&  $^{7}$Li&   $^{8}$Be         \\
\hline
0$s_{1/2}$            &  0.0575         &  0.128            \\
\hline 
0$p_{3/2}$            &  1.0491         &    1.076             \\
\hline 
1$s_{1/2}$            &  0.2437         &     0.232              \\
\hline 
0$p_{1/2}$            &  0.2111         &    0.214               \\
\hline 
0$d_{5/2}$            &    0.       &        0.272            \\
\hline 
\end{tabular}}
\end{table}

\begin{table}[h!]
{\begin{tabular}{|c|c|c|c|c|c|c|}
\cline{1-7} 
c) &  1$s_{1/2}$ &  0$d_{3/2}$ &  0$f_{7/2}$ &  1$p_{3/2}$ &  1$p_{1/2}$ &  0$f_{5/2}$         \\
\hline
$^{50}$Ca              &    0.063    &   0.0894    &  0.2041    &   0.9979     &    0.1628   &    0.177              \\
\hline 
\end{tabular}}
\end{table}

\begin{table}[h!]
{\begin{tabular}{|c|c|c|c|c|c|c|}
\cline{1-7} 
 d)&  0$h_{9/2}$ &  1$f_{7/2}$ &  0$i_{13/2}$ &  2$p_{3/2}$ &  1$f_{5/2}$ &  2$p_{1/2}$         \\
\hline
$^{206}$Pb              &    0.14    &   0.18    &  0.28    &   0.28     &    0.47   &    0.75              \\
\hline 
\end{tabular}}
\end{table}

\begin{table}[h!]
{\begin{tabular}{|c|c|c|c|}
\cline{1-4} 
 e)&  0$d_{5/2}$ &  1$s_{1/2}$ &  0$d_{3/2}$         \\
\hline
$^{18}$O              &  0.89  &  0.396   & 0.223          \\
\hline 
\end{tabular}
\begin{equation*}
 \left|^{12}\textrm{Be}(gs)\right.\rangle \left\{ \begin{array}{l l} |0\rangle + \alpha |(p_{1/2},s_{1/2})_{1^-} \otimes 1^-;0^+ \rangle + 
                                                             \beta  |(s_{1/2},d_{5/2})_{2^+} \otimes 2^+;0^+ \rangle +
                                                             \gamma |(p_{1/2},d_{5/2})_{3^-} \otimes 3^-;0^+ \rangle  \\
                                                 \alpha = 0.10, \beta = 0.35, \gamma=0.33,                         \\
                                                 |0\rangle = 0.37 |s^2_{1/2}(0)\rangle + 0.50|p^2_{1/2}(0)\rangle + 0.60 |d^2_{1/2}(0)\rangle
                              \end{array} \right.
\end{equation*}
\begin{equation*}
 \left|^{11}\textrm{Li}(gs)\right.\rangle \left\{ \begin{array}{l l} |0\rangle + \alpha |(p_{1/2},s_{1/2})_{1^-} \otimes 1^-;0^+ \rangle + 
                                                             \beta  |(s_{1/2},d_{5/2})_{2^+} \otimes 2^+;0^+ \rangle \\
                                                 \alpha = 0.7, \beta = 0.1                     \\
                                                 |0\rangle = 0.45 |s^2_{1/2}(0)\rangle + 0.55|p^2_{1/2}(0)\rangle + 0.04 |d^2_{5/2}(0)\rangle
                              \end{array} \right.
\end{equation*}}
\captionsetup{singlelinecheck=off,justification=raggedright}
\caption[Two-nucleon spectroscopic amplitudes]{(a) Two-nucleon spectroscopic amplitudes $<BCS(A)|P_{\nu}|BCS(A+2)>$ = $\sqrt{(2j_{\nu}+1)/2} \;  U_{\nu}(A)V_{\nu}(A+2)$, associated with 
two-nucleon pick-up reaction connecting the ground state (member of a pairing rotational band) of two superfluid Sn-nuclei
$(^{A+2}$Sn$(p,t)^A$Sn(gs)); (b) Two-nucleon spectroscopic amplitudes associated with the pair removal modes $|r\rangle$ of the $N_0=6$  closed shell systems
$^9$Li and $^{10}$Be
\begin{equation*}
 <r|P_{\nu}|N_0=6> = \left\{ 
\begin{array}{l l}
X_{rem}(\nu), \; \epsilon_{\nu} \leq \epsilon_F \\
Y_{rem}(\nu), \; \epsilon_{\nu} > \epsilon_F \\
\end{array} \right.
\end{equation*}
(c) Two-nucleon spectroscopic amplitude associated with the pair addition mode $|a \rangle$ of the closed shell system $^{48}$Ca,
\begin{equation*}
 <a|P^ \dagger_{\nu}|N_0> = \left\{ 
\begin{array}{l l}
X_{add}(\nu), \; \epsilon_{\nu}  >\epsilon_F \\
Y_{add}(\nu), \; \epsilon_{\nu}  \leq\epsilon_F \\
\end{array} \right.
\end{equation*}
(d) Same as (b) but for the case of the closed shell $N_0= 126$ nucleus $^{208}$Pb.
(e) Same as (c) but for the closed shell nucleus 
($N_0=8$) $^{16}$O. 
In the bottom part of the table, the ground state wavefunctions associated with the pair addition mode of the 
$N_0=6$ closed shell systems 
$^{10}$Be and $^9$Li are displayed. 
For details concerning these wavefunctions cf. references \cite{Barranco:01,Gori:04}.
}\label{table:1}
\end{table}

\begin{table}[h!]
{\begin{tabular}{|c|c|c|c|c|c|c|c|c|c|c|c|c|}
\cline{2-13} 
\multicolumn{1}{c|}{}& \multicolumn{12}{|c|}{$^{A}$Sn($p,t)^{A-2}$Sn}           \\
\cline{2-13} 
\multicolumn{1}{c|}{} & $V$ & $W$ &  $V_{so}$ &  $W_d$ &  $r_1$ &  $a_1$ &  $r_2$ &  $a_2$ &  $r_3$ &  $a_3$ &  $r_4$ &  $a_4$            \\
\hline 
$p$,\;$^A$Sn$\,^{a)}$ & $50$ & $5$ &  $3$ &  $6$ &  $1.35$ &  $0.65$ &  $1.2$ &  $0.5$ &  $1.25$ &  $0.7$ &  $1.3$ &  $0.6$ \\
\hline 
$d$,\;$^{A-1}$Sn$\,^{b)}$ & $78.53$ & $12$ &  $3.62$ &  $10.5$ &  $1.1$ &  $0.6$ &  $1.3$ &  $0.5$ &  $0.97$ &  $0.9$ &  $1.3$ &  $0.61$ \\
\hline 
$t$,\;$^{A-2}$Sn$\,^{a)}$ & $176$ & $20$ &  $8$ &  $8$ &  $1.14$ &  $0.6$ &  $1.3$ &  $0.5$ &  $1.1$ &  $0.8$ &  $1.3$ &  $0.6$ \\
\hline
\multicolumn{1}{c|}{}& \multicolumn{12}{|c|}{$^{7}$Li($t,p)^{9}$Li}           \\
\cline{2-13} 
\multicolumn{1}{c|}{} & $V$ & $W$ &  $V_{so}$ &  $W_d$ &  $r_1$ &  $a_1$ &  $r_2$ &  $a_2$ &  $r_3$ &  $a_3$ &  $r_4$ &  $a_4$            \\
\hline 
$t$,\;$^{7}$Li$\,^{c)}$ & $150.93$ & $12.74$ &  $1.9$ &  $20.6$ &  $1.04$ &  $0.72$ &  $1.23$ &  $1.15$ &  $0.54$ &  $0.25$ &  $1.01$ &  $0.83$ \\
\hline 
$d$,\;$^{8}$Li$\,^{b)}$ & $89.1$ & $0$ &  $3.56$ &  $4.8$ &  $1.15$ &  $0.74$ &   &   &  $0.97$ &  $1.01$ &  $1.41$ &  $0.65$ \\
\hline 
$p$,\;$^{9}$Li$\,^{d)}$ & $56.06$ & $0$ &  $4.42$ &  $5.32$ &  $1.11$ &  $0.68$ &   &   &  $0.87$ &  $0.59$ &  $1.31$ &  $0.52$ \\
\hline 
\multicolumn{1}{c|}{}& \multicolumn{12}{|c|}{$^{11}$Li($p,t)^{9}$Li}           \\
\cline{2-13} 
\multicolumn{1}{c|}{} & $V$ & $W$ &  $V_{so}$ &  $W_d$ &  $r_1$ &  $a_1$ &  $r_2$ &  $a_2$ &  $r_3$ &  $a_3$ &  $r_4$ &  $a_4$            \\
\hline 
$p$,\;$^{11}$Li$\,^{d)}$ & $63.62$ & $0.33$ &  $5.69$ &  $8.9$ &  $1.12$ &  $0.68$ &  $1.12$ &  $0.52$ &  $0.89$ &  $0.59$ &  $1.31$ &  $0.52$ \\
\hline 
$d$,\;$^{10}$Li$\,^{b)}$ & $90.76$ & $1.6$ &  $3.56$ &  $10.58$ &  $1.15$ &  $0.75$ &  $1.35$ &  $0.64$ &  $0.97$ &  $1.01$ &  $1.4$ &  $0.66$ \\
\hline 
$t$,\;$^{9}$Li$\,^{c)}$ & $152.47$ & $12.59$ &  $1.9$ &  $12.08$ &  $1.04$ &  $0.72$ &  $1.23$ &  $0.72$ &  $0.53$ &  $0.24$ &  $1.03$ &  $0.83$ \\
\hline 
\multicolumn{1}{c|}{}& \multicolumn{12}{|c|}{$^{10}$Be($t,p)^{12}$Be}           \\
\cline{2-13} 
\multicolumn{1}{c|}{} & $V$ & $W$ &  $V_{so}$ &  $W_d$ &  $r_1$ &  $a_1$ &  $r_2$ &  $a_2$ &  $r_3$ &  $a_3$ &  $r_4$ &  $a_4$            \\
\hline 
$t$,\;$^{10}$Be$\,^{e)}$ & $195$ & $18.9$ &  $0$ &  $0$ &  $1.29$ &  $0.58$ &  $1.37$ &  $0.96$ &   &   &   &  \\
\hline 
$d$,\;$^{11}$Be$\,^{b)}$ & $91.37$ & $1.51$ &  $3.58$ &  $10.63$ &  $1.15$ &  $0.75$ &  $1.35$ &  $0.63$ &  $0.97$ &  $1.01$ &  $1.4$ &  $0.67$ \\
\hline 
$p$,\;$^{12}$Be$\,^{e)}$ & $90$ & $0$ &  $5.5$ &  $8.55$ &  $1.13$ &  $0.57$ &   &   &  $1.13$ &  $0.57$ &  $1.13$ &  $0.5$ \\
\hline 
\multicolumn{1}{c|}{}& \multicolumn{12}{|c|}{$^{48}$Ca($t,p)^{50}$Ca}           \\
\cline{2-13} 
\multicolumn{1}{c|}{} & $V$ & $W$ &  $V_{so}$ &  $W_d$ &  $r_1$ &  $a_1$ &  $r_2$ &  $a_2$ &  $r_3$ &  $a_3$ &  $r_4$ &  $a_4$            \\
\hline 
$t$,\;$^{48}$Ca$\,^{f)}$ & $144$ & $20$ &  $0$ &  $0$ &  $1.24$ &  $0.68$ &  $1.35$ &  $0.84$ &  &   &   &   \\
\hline 
$d$,\;$^{49}$Ca$\,^{b)}$ & $85$ & $5$ &  $3.$ &  $13$ &  $1.15$ &  $0.9$ &  $1.33$ &  $0.74$ &  $0.8$ &  $0.74$ &  $1.3$ &  $0.8$ \\
\hline 
$p$,\;$^{50}$Ca$\,^{f)}$ & $49$ & $10$ &  $0$ &  $10.5$ &  $1.25$ &  $0.65$ &  $1.25$ &  $0.47$ &   &   &  $1.25$ &  $0.47$ \\
\hline 
\multicolumn{1}{c|}{}& \multicolumn{12}{|c|}{$^{206}$Pb($t,p)^{208}$Pb}           \\
\cline{2-13} 
\multicolumn{1}{c|}{} & $V$ & $W$ &  $V_{so}$ &  $W_d$ &  $r_1$ &  $a_1$ &  $r_2$ &  $a_2$ &  $r_3$ &  $a_3$ &  $r_4$ &  $a_4$            \\
\hline 
$t$,\;$^{206}$Pb $\,^{g)}$& $200$ & $0$ &  $0$ &  $10$ &  $1.4$ &  $0.6$ &   &   &  &   & 1.4  & 0.6  \\
\hline 
$d$,\;$^{207}$Pb$\,^{b)}$ & $97.78$ & $0$ &  $3.56$ &  $8$ &  $1.15$ &  $0.79$ &   &   &  $0.97$ &  $1.01$ &  $1.36$ &  $0.9$ \\
\hline 
$p$,\;$^{208}$Pb $\,^{g)}$& $55$ & $0$ &  $0$ &  $11.5$ &  $1.25$ &  $0.65$ &   &   &   &   &  $1.25$ &  $0.47$ \\
\hline 
\multicolumn{1}{c|}{}& \multicolumn{12}{|c|}{$^{208}$Pb($^{16}$O,$^{18}$O)$^{206}$Pb}           \\
\cline{2-13} 
\multicolumn{1}{c|}{} & $V$ & $W$ &  $V_{so}$ &  $W_d$ &  $r_1$ &  $a_1$ &  $r_2$ &  $a_2$ &  $r_3$ &  $a_3$ &  $r_4$ &  $a_4$            \\
\hline 
$^{16}$O,\;$^{208}$Pb$\,^{h)}$ & $100$ & $65.4$ &  $0$ &  $0$ &  $1.26$ &  $0.45$ & $1.26$  & 0.45  &  &   &   &   \\
\hline 
$^{17}$O,\;$^{207}$Pb$\,^{h)}$ & $100$ & $65.4$ &  $0$ &  $0$ &  $1.26$ &  $0.45$ & $1.26$  & 0.45  &  &   &   & \\
\hline 
$^{18}$O,\;$^{206}$Pb$\,^{h)}$ & $65$ & $45$ &  $0$ &  $0$ &  $1.35$ &  $0.34$ & 1.34  & 0.33  &   &   &   &   \\
\hline 
  \end{tabular}}
    \captionsetup{singlelinecheck=off,justification=raggedright}
   \caption{\protect Optical potentials used in the calculation of the absolute value of the two--nucleon transfer differential cross sections (see Figs. 5,6 and Table 3). The quantities $V,W,V_{SO},W_d$ are in MeV while the remaining in fm.\\
   The nuclear term of the optical potential was chosen to have the form
   $U(r)=-Vf_1(r)-iWf_2(r)-4iW_d\;g_3(r)-\left(\frac{\hbar}{m_\pi c}\right)^2 V_{so}\frac{g_4(r)}{a_4r}\mathbf{l\cdot s},$
   with $f_i(r)=\frac{1}{1+e^{(r-R_i)/a_i}};\quad g_i(r)=\frac{e^{(r-R_i)/a_i}}{\left(1+e^{(r-R_i)/a_i}\right)^2},$
   and $m_\pi$ the pion mass. For $(p,t)$ and $(t,p)$ reactions $R_i=r_iA^{1/3}$, where $A$ is the mass number of the heavy nucleus in the corresponding channel, while for reactions involving heavy nuclei of mass numbers $A_a, A_A$ $R_i=r_i(A_a^{1/3}+A_A^{1/3})$. The Coulomb term is taken to be the electrostatic potential generated by an uniformly charged sphere of radius $R_1$.\\
   ${}^{a)}$ P. Guazzoni, L. Zetta, et al., Phys. Rev. \textbf{C 74}, 054605 (2006).\newline
   ${}^{b)}$ Haixia An and Chonghai Cai, Phys. Rev. \textbf{C 73}, 054605 (2006). \newline
   ${}^{c)}$ Xiaohua Li, Chuntian Liang and Chonghai Cai,  Nucl. Phys. \textbf{A 789} 103 (2007).\newline
   ${}^{d)}$  A.J. Koning and J.P. Delaroche  Nucl. Phys. \textbf{A 713} 231 (2003).\newline
   ${}^{e)}$ H.T. Fortune, G.B. Liu and D.E. Alburger, Phys. Rev. \textbf{C 50}, (1994) 1355.\newline
   ${}^{f)}$ B. Bayman, Nucl. Phys. \textbf{A 168} 1 (1971).\newline
   ${}^{g)}$ R. A. Broglia and C. Riedel Nucl. Phys. \textbf{A 92} 145 (1967).\newline
   ${}^{h)}$  B. Bayman and J. Chen, Phys. Rev. \textbf{C 26} 1509 (1982).} 
\label{table:3}
\end{table}

\begin{table}[h!]
{  \begin{tabular}{|c|c|c|c|}
\cline{2-4} 
\multicolumn{1}{c|}{}                                                                & \multicolumn{3}{|c|}{$\sigma($gs$\rightarrow$f)}              \\
\cline{2-4}
\multicolumn{1}{c|}{}                                                                & f        & Theory ${}^{a)}$ ${}^{b)}$          & Experiment${}^{f-n)}$                        \\
\hline 
$^{7}$Li($t,p$)$^{9}$Li                                                              & gs   & 14.3 ${}^{c)}$  & $14.7 \pm 4.4$ ${}^{c,i)}\quad[9.4^\circ<\theta<108.7^\circ]$                    \\
\hline
\multirow{2}{*}{$^1$H($^{11}$Li,$^9$Li)$^3$H} & gs       & 6.1 ${}^{c)}$  & $5.7 \pm 0.9 $ ${}^{c,b)}\quad[20^\circ<\theta<154.5^\circ]$           \\
\cline{2-4}
                                                                                     & $1/2^{-}$& 0.7 ${}^{c)}$  & $1.0 \pm 0.36$ ${}^{c,b)}\quad[30^\circ<\theta<100^\circ]$          \\
\hline
$^{10}$Be($t,p$)$^{12}$Be                                                            & gs    & 2.3 ${}^{c)}$  & $1.9 \pm 0.57 $ ${}^{c,j)}\quad[4.4^\circ<\theta<57.4^\circ]$             \\
\hline
$^{48}$Ca($t,p$)$^{50}$Ca                                                          & gs       & 0.55 ${}^{c)}$ & $0.56 \pm 0.17$ ${}^{c,m)}\quad[4.5^\circ<\theta<174^\circ]$         \\
\hline
\hline 
$^{112}$Sn($p,t$)$^{110}$Sn, $E_{CM}=26$ MeV                                            & gs       & 1301 ${}^{d)}$           & $1309 \pm 200 (\pm 14)$ ${}^{d,g)}\quad[6^\circ<\theta<62.2^\circ] $\\
\hline 
$^{114}$Sn($p,t$)$^{112}$Sn, $E_{CM}=22$ MeV                                            & gs       & 1508  ${}^{d)}$           & $1519 \pm 456 (\pm 16.2)$ ${}^{d,g)}\quad[7.64^\circ<\theta<62.24^\circ]$\\
\hline 
$^{116}$Sn($p,t$)$^{114}$Sn, $E_{CM}=26$ MeV                                            & gs       & 2078 ${}^{d)}$           & $2492 \pm 374 (\pm 32)$ ${}^{d,g)}\quad[4^\circ<\theta<70^\circ]$\\
\hline 
$^{118}$Sn($p,t$)$^{116}$Sn, $E_{CM}=24.4$ MeV                                            & gs       & 1304 ${}^{d)}$           & $1345 \pm 202 (\pm 24)$ ${}^{d,g)}\quad[7.63^\circ<\theta<59.6^\circ]$\\
\hline 
$^{120}$Sn($p,t$)$^{118}$Sn, $E_{CM}=21$ MeV                                            & gs       & 2190 ${}^{d)}$           & $2250 \pm 338 (\pm 14)$ ${}^{d,g)}\quad[7.6^\circ<\theta<69.7^\circ]$\\
\hline
${}^{122}\textrm{Sn}(p,t){}^{120}$Sn, $E_{CM}=26$ MeV                                   & gs       & 2466 ${}^{d)}$           & $2505 \pm 376 (\pm 18)$ ${}^{d,g)}\quad[6^\circ<\theta<62.2^\circ]$\\
\hline 
$^{124}$Sn($p,t$)$^{122}$Sn, $E_{CM}=25$ MeV                                            & gs       & 838  ${}^{d)}$           & $958 \pm 144 (\pm 15)$ ${}^{d,g)}\quad[4^\circ<\theta<57^\circ]$\\
\hline
\hline 
$^{112}$Sn($p,t$)$^{110}$Sn, $E_p=40$ MeV                                             & gs       & 3349 ${}^{e)}$ & $3715 \pm 1114$ ${}^{e,h)}$ \\
\hline
$^{114}$Sn($p,t$)$^{112}$Sn, $E_p=40$ MeV                                             & gs       & 3790 ${}^{e)}$ & $3776 \pm 1132$ ${}^{e,h)}$ \\
\hline
$^{116}$Sn($p,t$)$^{114}$Sn, $E_p=40$ MeV                                             & gs       & 3085 ${}^{e)}$ & $3135 \pm 940$ ${}^{e,h)}$ \\
\hline
$^{118}$Sn($p,t$)$^{116}$Sn, $E_p=40$ MeV                                             & gs       & 2563 ${}^{e)}$ & $2294 \pm 668$ ${}^{e,h)}$ \\
\hline
$^{120}$Sn($p,t$)$^{118}$Sn, $E_p=40$ MeV                                             & gs       & 3224 ${}^{e)}$ & $3024 \pm 907$ ${}^{e,h)}$ \\
\hline
$^{122}$Sn($p,t$)$^{120}$Sn, $E_p=40$ MeV                                             & gs       & 2339 ${}^{e)}$ & $2907 \pm 872$ ${}^{e,h)}$ \\
\hline
$^{124}$Sn($p,t$)$^{122}$Sn, $E_p=40$ MeV                                             & gs       & 1954 ${}^{e)}$ & $2558 \pm 767$ ${}^{e,h)}$ \\
\hline 
\hline
$^{206}$Pb($t,p$)$^{208}$Pb                                                          & gs       & 0.52 ${}^{c)}$ & $0.68 \pm 0.21$ ${}^{c,k)}\quad[4.5^\circ<\theta<176.5^\circ]$         \\
\hline
$^{208}$Pb($^{16}$O,$^{18}$O)$^{206}$Pb                                              & gs       & 0.80 ${}^{c)}$  & $0.76 \pm 0.18$ ${}^{c,f)}\quad[84.6^\circ<\theta<157.3^\circ]$         \\
\hline 
  \end{tabular} }
    \captionsetup{singlelinecheck=off,justification=raggedright}
 \caption{\protect\\Absolute value of two--nucleon transfer cross sections. 
   The number in parenthesis (last column) corresponds to\\ the statistical errors. \newline   
   ${}^{a)}$ G. Potel et al., Phys. Rev. Lett. \textbf{107}, (2011) 092501.\newline   
   ${}^{b)}$ G. Potel et al., Phys. Rev. Lett. \textbf{105}, (2010) 172502. \newline   
   ${}^{c)}$ mb \newline
   ${}^{d)}$ $\mu$b \newline
   ${}^{e)}$ $\mu$b/sr ($\sum_{i=1}^{N}(d\sigma/d\Omega)$; differential cross section summed over the few, $N=3-7$ experimental points).\newline
   ${}^{f)}$  B. Bayman and J. Chen, Phys. Rev. \textbf{C 26} (1982) 1509 and refs. therein.\newline
   ${}^{g)}$ P. Guazzoni, L. Zetta, et al., Phys. Rev. \textbf{C 60}, 054603 (1999).\newline 
    P. Guazzoni, L. Zetta, et al., Phys. Rev. \textbf{C 69}, 024619 (2004).\newline    
    P. Guazzoni, L. Zetta, et al., Phys. Rev. \textbf{C 74}, 054605 (2006).\newline    
    P. Guazzoni, L. Zetta, et al., Phys. Rev. \textbf{C 83}, 044614 (2011).\newline    
    P. Guazzoni, L. Zetta, et al., Phys. Rev. \textbf{C 78}, 064608 (2008).\newline    
    P. Guazzoni, L. Zetta, et al., Phys. Rev. \textbf{C 85}, 054609 (2012).\newline
   ${}^{h)}$ G. Bassani et al. Phys. Rev. \textbf{139}, (1965)B830.\newline
   ${}^{i)}$ P.G. Young and R.H. Stokes, Phys. Rev. \textbf{C 4}, (1971) 1597.\newline    
   ${}^{j)}$ H.T. Fortune, G.B. Liu and D.E. Alburger, Phys. Rev. \textbf{C 50}, (1994) 1355.\newline
   ${}^{k)}$ J.H. Bjerregaard et al., Nucl. Phys. \textbf{89}, (1966) 337.\newline
   ${}^{m)}$ J. H. Bjerregaard et al., Nucl. Phys. \textbf{A 103}, (1967) 33.\newline
   ${}^{n)}$ Tanihata et al., Phys. Rev. Lett. \textbf{100}, (2008) 192502.}
\label{table:4}
\end{table}
\begin{figure}[h!]
	\begin{center}
		\includegraphics[width=0.75\textwidth]{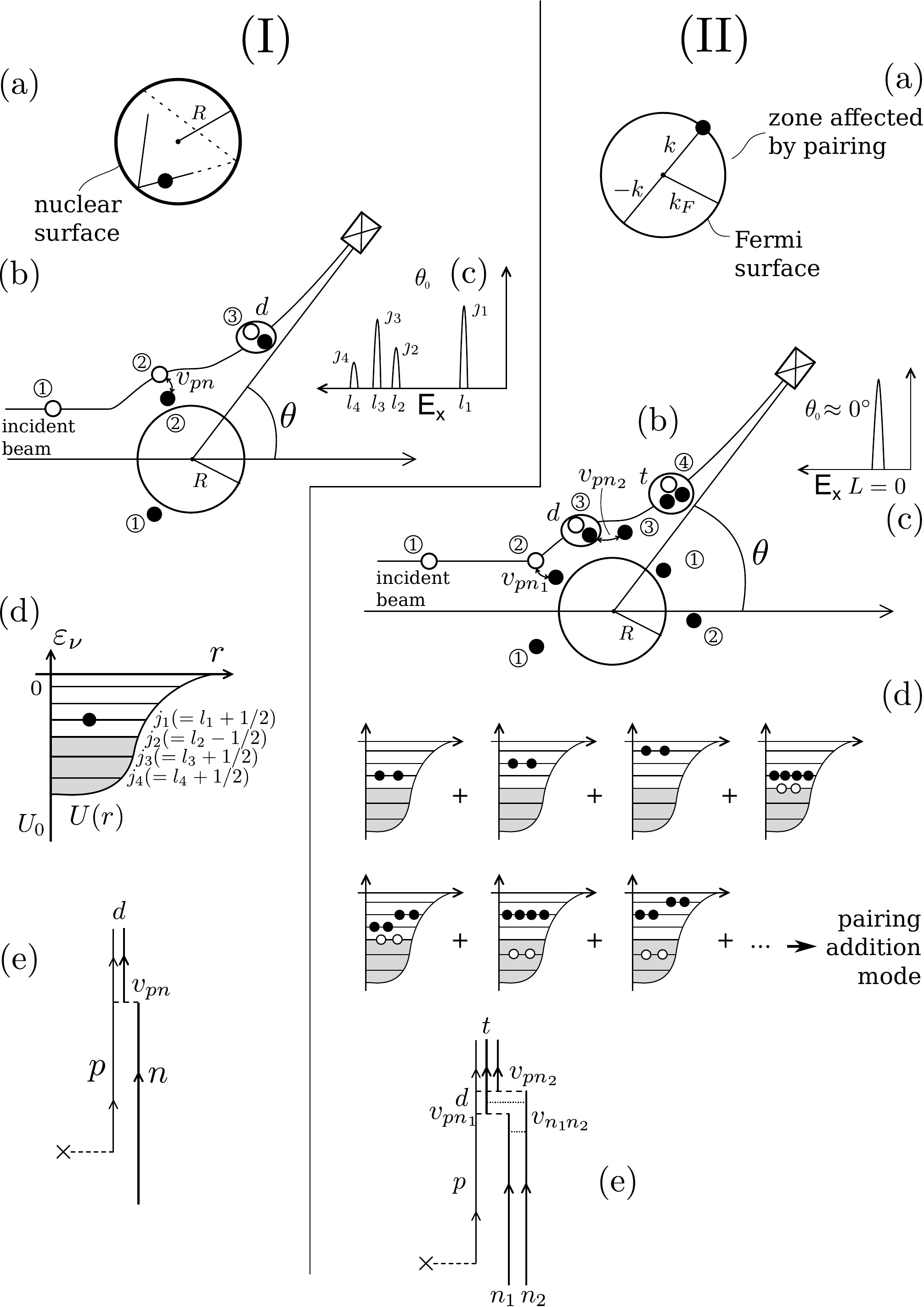}
	\end{center}
	\captionsetup{singlelinecheck=off,justification=raggedright}
	\caption{Schematic representation of single-particle and of pairing vibration elementary modes of nuclear excitation
 (I): {\bf (a)} Cartoon (semiclassical) representation of a nucleon moving in the mean field produced by the other nucleons; {\bf (b)} the specific probe to test the validity of the nuclear independent-particle model is to be found in one-particle transfer reactions (circled numerals 1,2... indicating different milestones of the transfer process), as testified by {\bf (c)} the sharp peaks observed in e.g. $N_0(p,d)N_0-1$ neutron pick-up reactions ($N_0$: neutron magic number), peaks which provide information concerning: (d) the single-particle levels of the mean field potential $U(r) = \int \textrm{d}^3 r \rho(r)v(|r-r'|)$, (e) the interaction $v$ acting in the process in which a neutron moves from $N_0$ to the deuteron.
(II): {\bf (a)} Schematic representation of the fact that pairing correlations in nuclei are specifically probed by two-nucleon transfer processes. Cooper's model associated with a pair of correlated fermions moving in time reversal states $(k \uparrow, -k \downarrow)$ on top of the Fermi surface. {\bf (b)} In keeping with the fact that the correlation length is larger than nuclear dimensions (see Fig. 2) and that $U(R_0) >> G (\approx 25/A$ MeV), the natural mechanism for pair transfer is successive transfer (circled numerals 1,2... indicating different milestones of the transfer process), as exemplified in the case of the reaction $(A_0+2)(p,t)A_0(gs)$, and testified by {\bf (c)} the  $L=0$ levels observed experimentally (see Fig. 3), in keeping with the fact that the  mode studied is: {\bf (d)} a linear combination of $2p, 4p-2h,$ etc. excitations. {\bf (e)} Because the Cooper pair correlation is weak, in the main contribution to the two-nucleon transfer process the interaction acts twice (see also Fig. 7).}
\label{fig:8}
\end{figure}

\begin{figure}[h!]
	\begin{center}
		\includegraphics[width=0.78\textwidth]{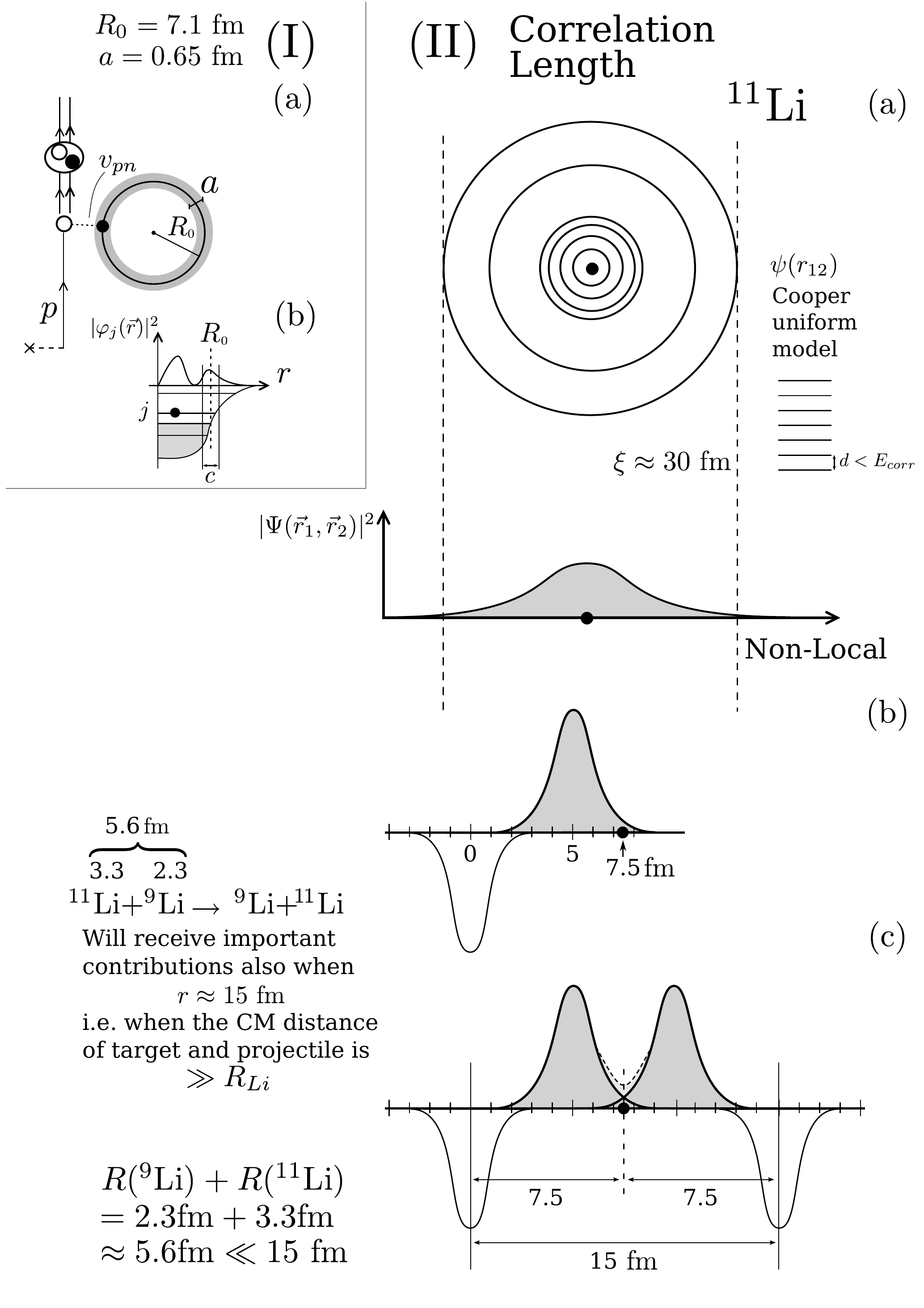}
	\end{center}
	\captionsetup{singlelinecheck=off,justification=raggedright}
	\caption{ 
Schematic representation of the modulus square of the formfactors associated with one-particle transfer and with Cooper-pair tunneling (two-particle transfer).
I) Cartoon emphasizing the local aspect of (a) one--particle pick-up, where the process $(N_0+1)(p,d)N_0$ is mediated by the interaction $v_{np}$, and the formfactor $\varphi_j(r)$; (b) the square modulus $|\varphi_j (\vec r)|^2$ is the probability of finding a neutron moving above the Fermi surface in the mean field of the closed shell system $N_0$.
II) Making use of an uniform set of single-particle levels and allowing  neutrons to interact in terms of a schematic (density--dependent) pairing force whose strength is adjusted so that $E_{corr} \approx 400$ keV, one obtains  (a) Cooper pair wavefunction whose modulus squared testifies to the fact that the two-neutrons are correlated over distances of $\approx 30$ fm. (b) Introducing an external field parametrized as a standard Saxon-Woods potential (see e.g. Bohr and Mottelson, Nuclear Structure Vol I) with $N=8$ and $Z=3$ the above probability density is strongly modified, a modification which is repeated as shown in (c) by switching on another external field (standard Saxon-Woods potential) associated with $N=6$ and $Z=3$, placed at a distance of 15 fm from the first one. The very non-local aspect of the resulting single Cooper pair density probability, shared by two $^{9}$Li cores, can be probed in e.g. the reaction $^{11}$Li+$^{9}$Li $\rightarrow { }^{9}$Li + $^{11}$Li.
}
\label{fig:9}
\end{figure}

\begin{figure}[h!]
	\begin{center}
		\includegraphics[width=0.88\textwidth]{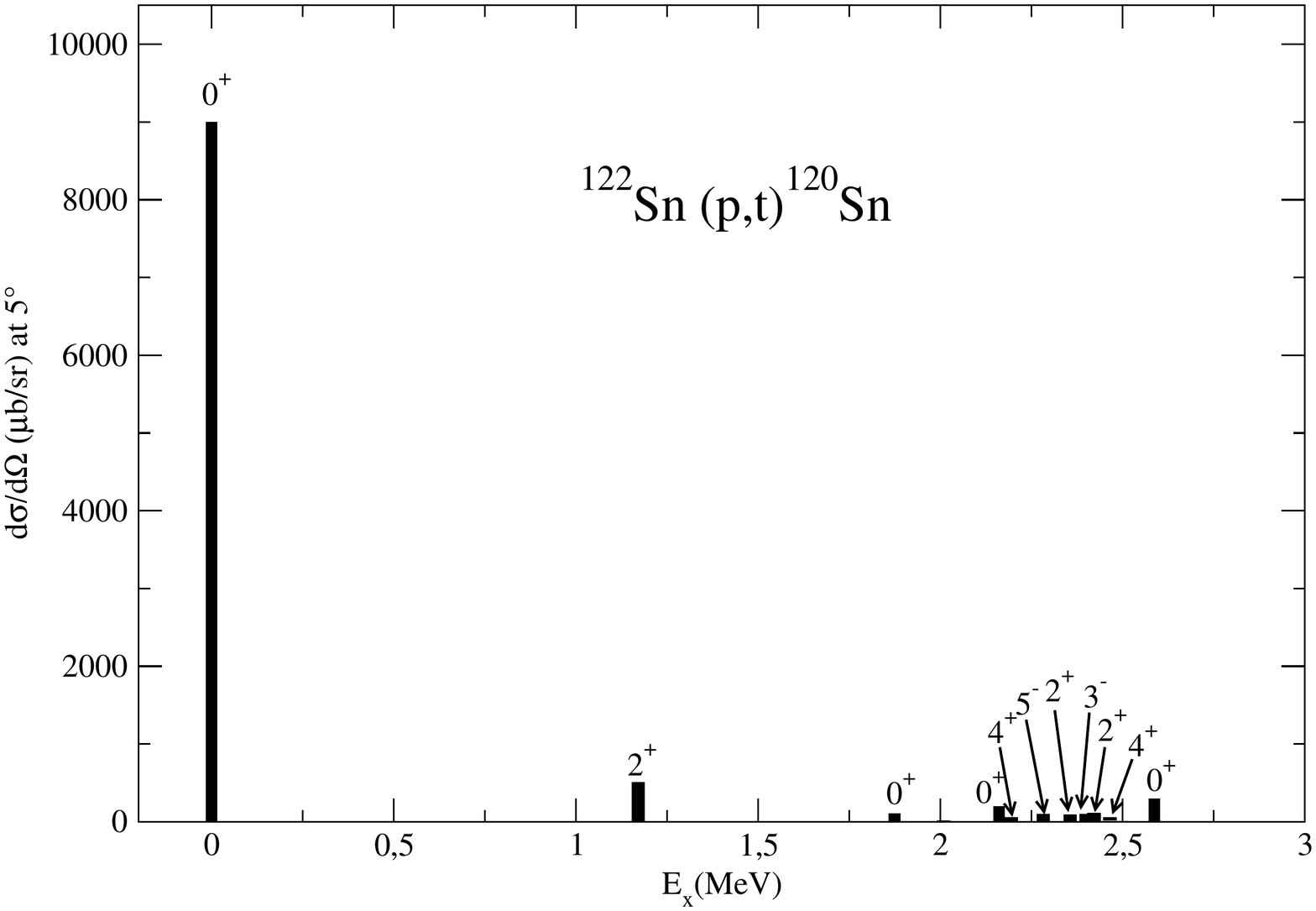}
	\end{center}
	\captionsetup{singlelinecheck=off,justification=raggedright}
	\caption{Excitation function associated with the reaction $^{122}$Sn$(p,t)^{120}$Sn$(J^{\pi})$. The absolute experimental value (see ref. $g)$ of  Table 3) of $\left. d \sigma(J^\pi)/d \Omega \right|_{5^{\circ}}$ is given as a function of the excitation energy $E_x$.}
\label{fig:4}
\end{figure}

\begin{figure}[h!]
	\begin{center}
		\includegraphics[width=0.98\textwidth]{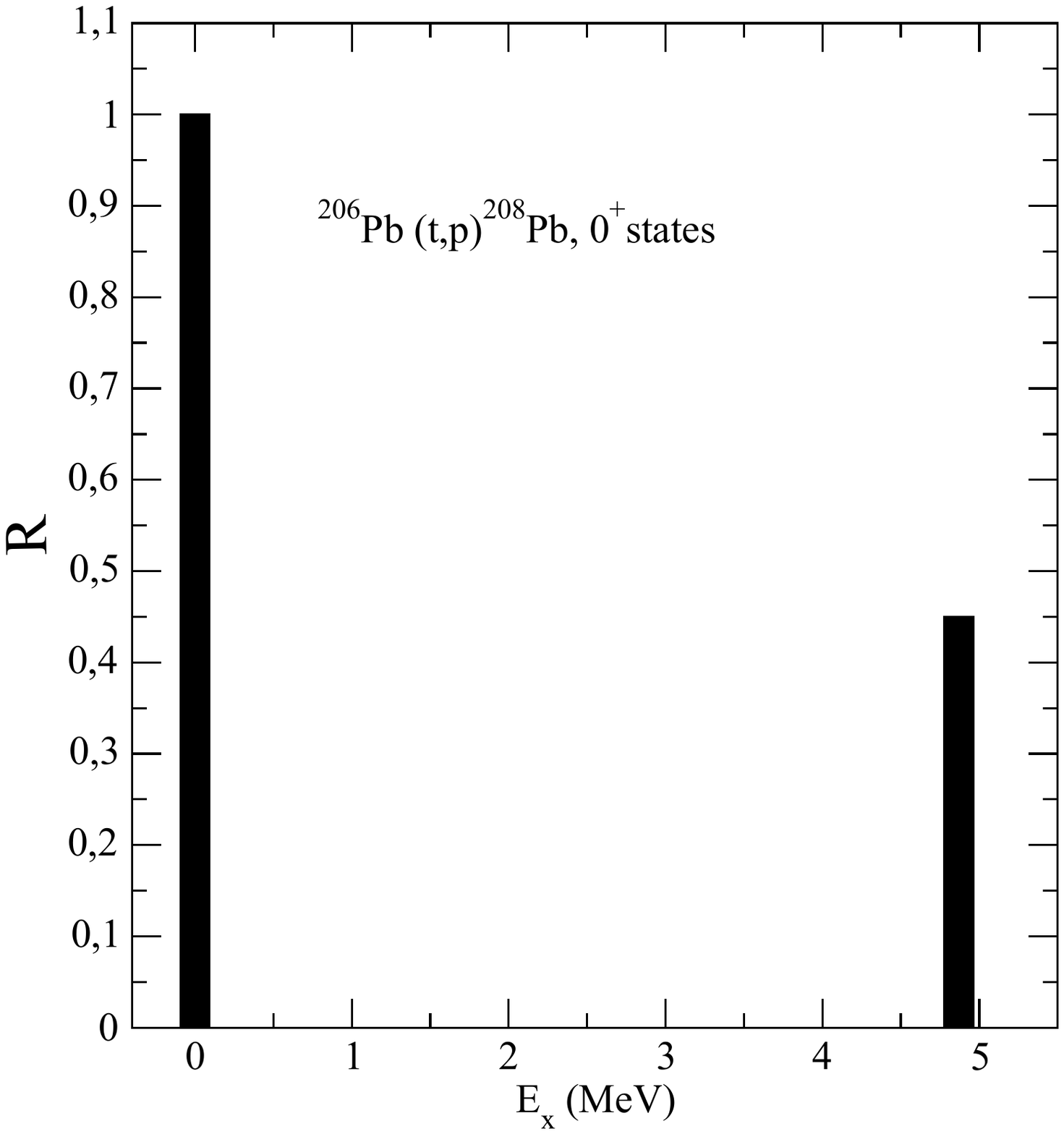}
	\end{center}
	\captionsetup{singlelinecheck=off,justification=raggedright}
	\caption{Excitation function associated with the reaction $^{206}$Pb$(t,p)^{208}$Pb$(0^{+})$. The ratio $R$ of the integrated absolute differential cross section associated with excited $0^+$ states below 5 MeV to the gs $\rightarrow$ gs absolute cross section is given as a function of the excitation energy $E_x$ (see ref. $k)$ of Table 3).}
\label{fig:6}
\end{figure}


\begin{figure}[h!]
	\begin{center}
		\includegraphics[width=0.88\textwidth]{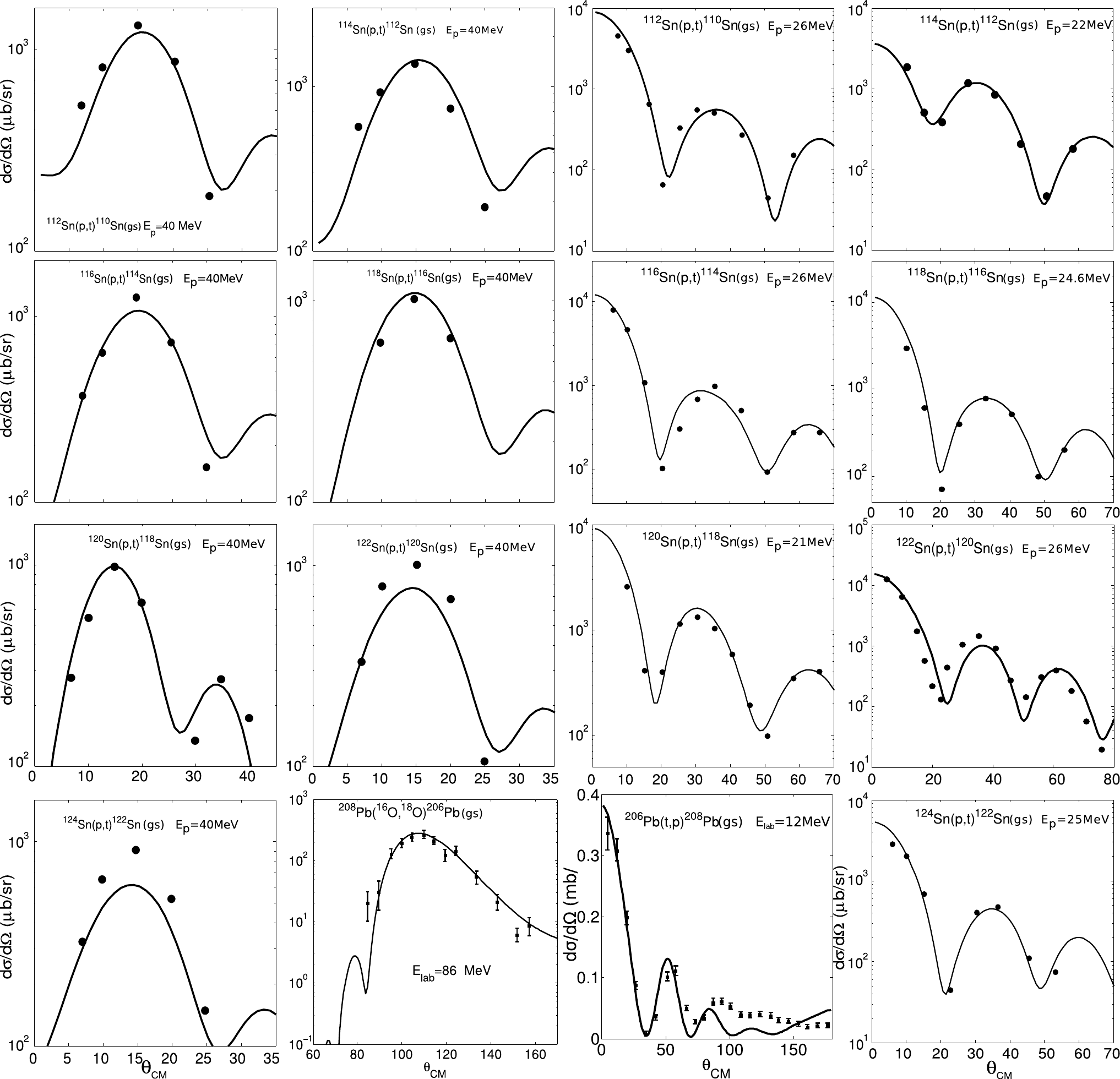}
	\end{center}
	\captionsetup{singlelinecheck=off,justification=raggedright}
	\caption{Absolute cross sections associated with a number of two--particle transfer reactions. Making use of the two-nucleon spectroscopic amplitudes and of the optical potentials collected in Tables 1  and 2, and of a two--nucleon transfer software developed by Gregory Potel \cite{Potel:unp} within the framework of second--order DWBA taking into account non-orthogonality corrections. The absolute differential cross sections associated with a number of reactions were calculated and are displayed (continuous curves) in comparison with the experimental data (cf. also Table 3).}
\label{fig:12}
\end{figure}

\begin{figure}[h!]
	\begin{center}
		\includegraphics[width=0.88\textwidth]{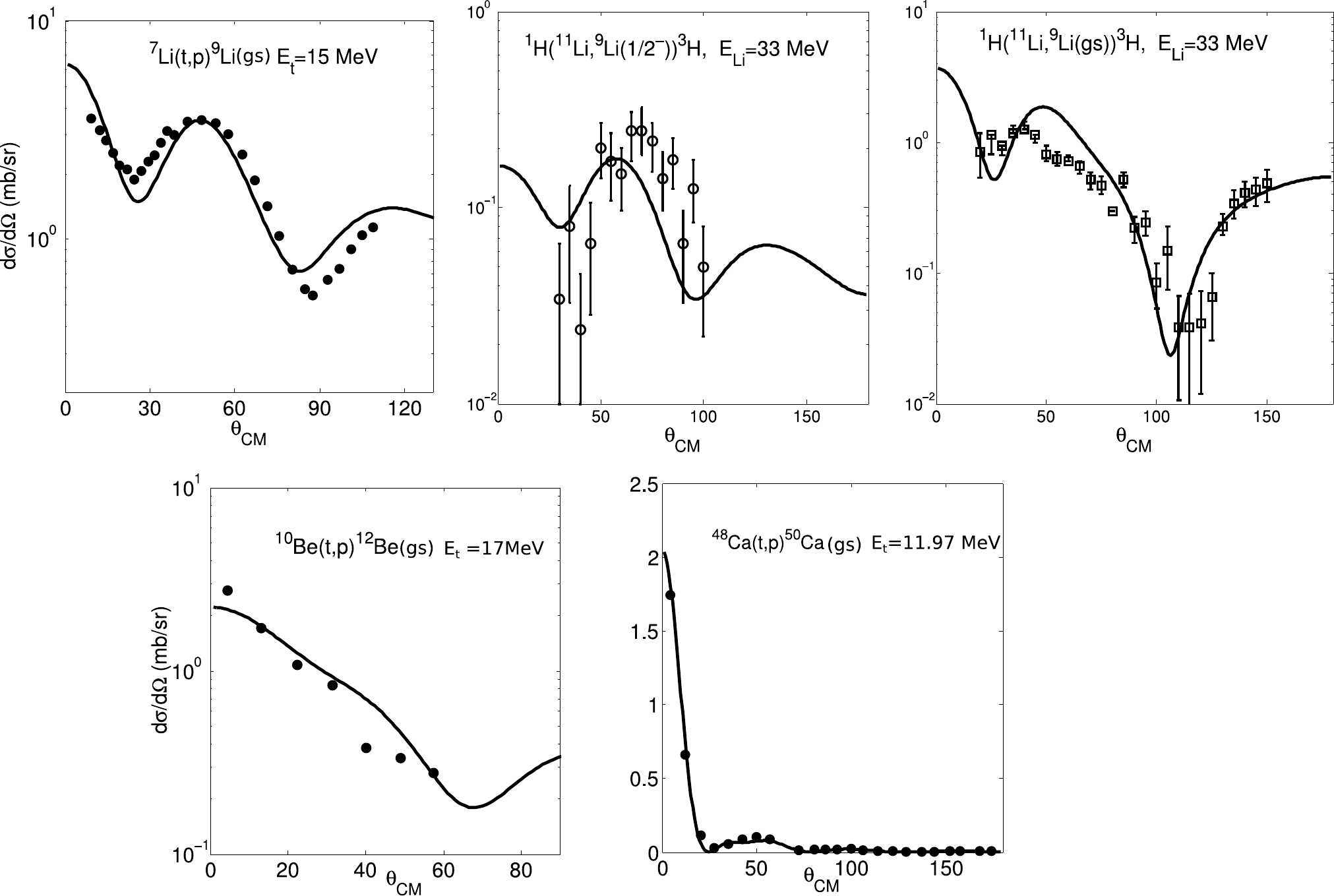}
	\end{center}
	\captionsetup{singlelinecheck=off,justification=raggedright}
	\caption{Same as Fig. 5 but for transfer processes involving medium light nuclei (see also Table 3 ).}
\label{fig:13}
\end{figure}
\begin{figure}[h!]
	\begin{center}
		\includegraphics[width=0.75\textwidth]{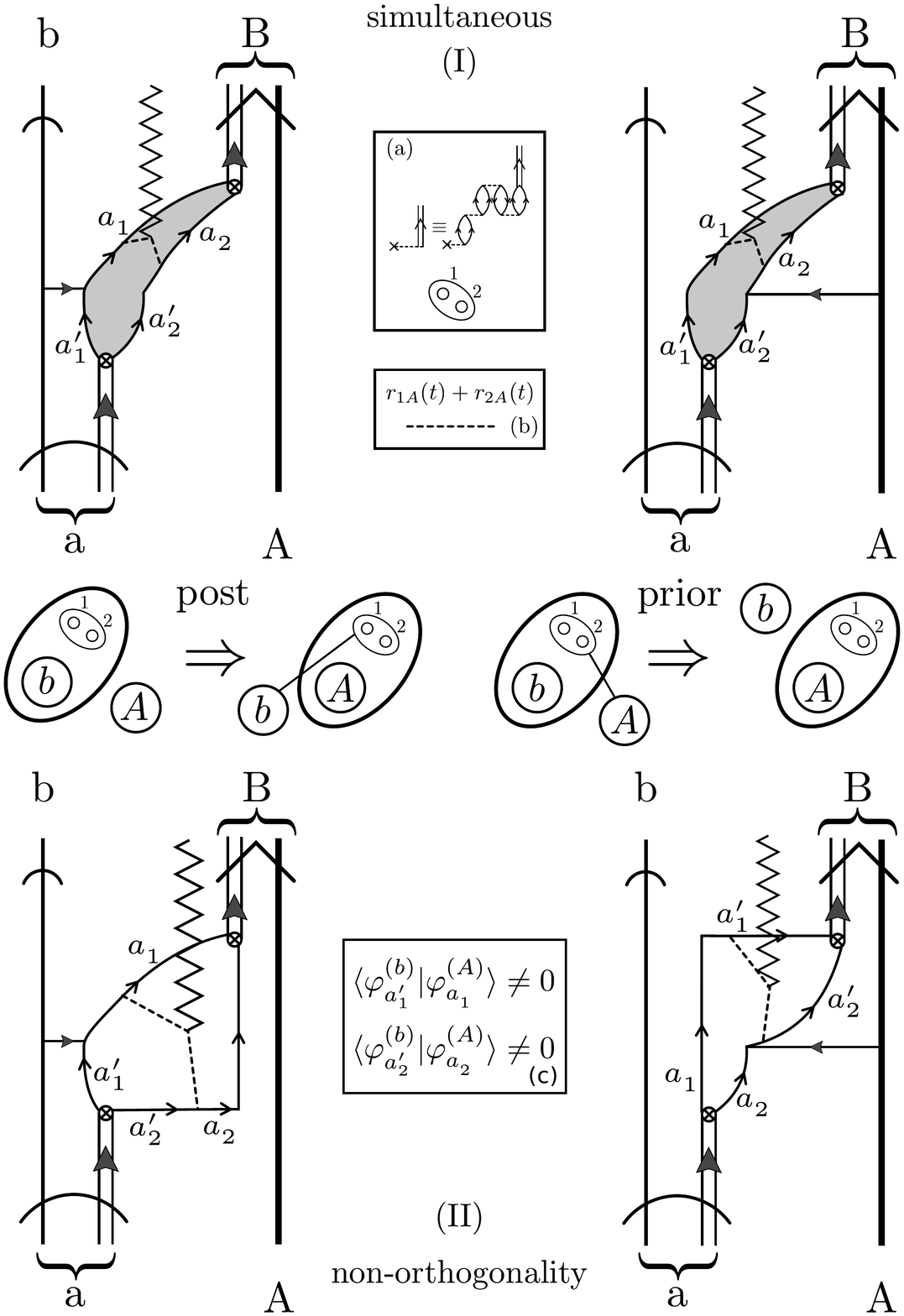}
	\end{center}
\end{figure}
\begin{figure}[h!]
	\begin{center}
		\includegraphics[width=0.75\textwidth]{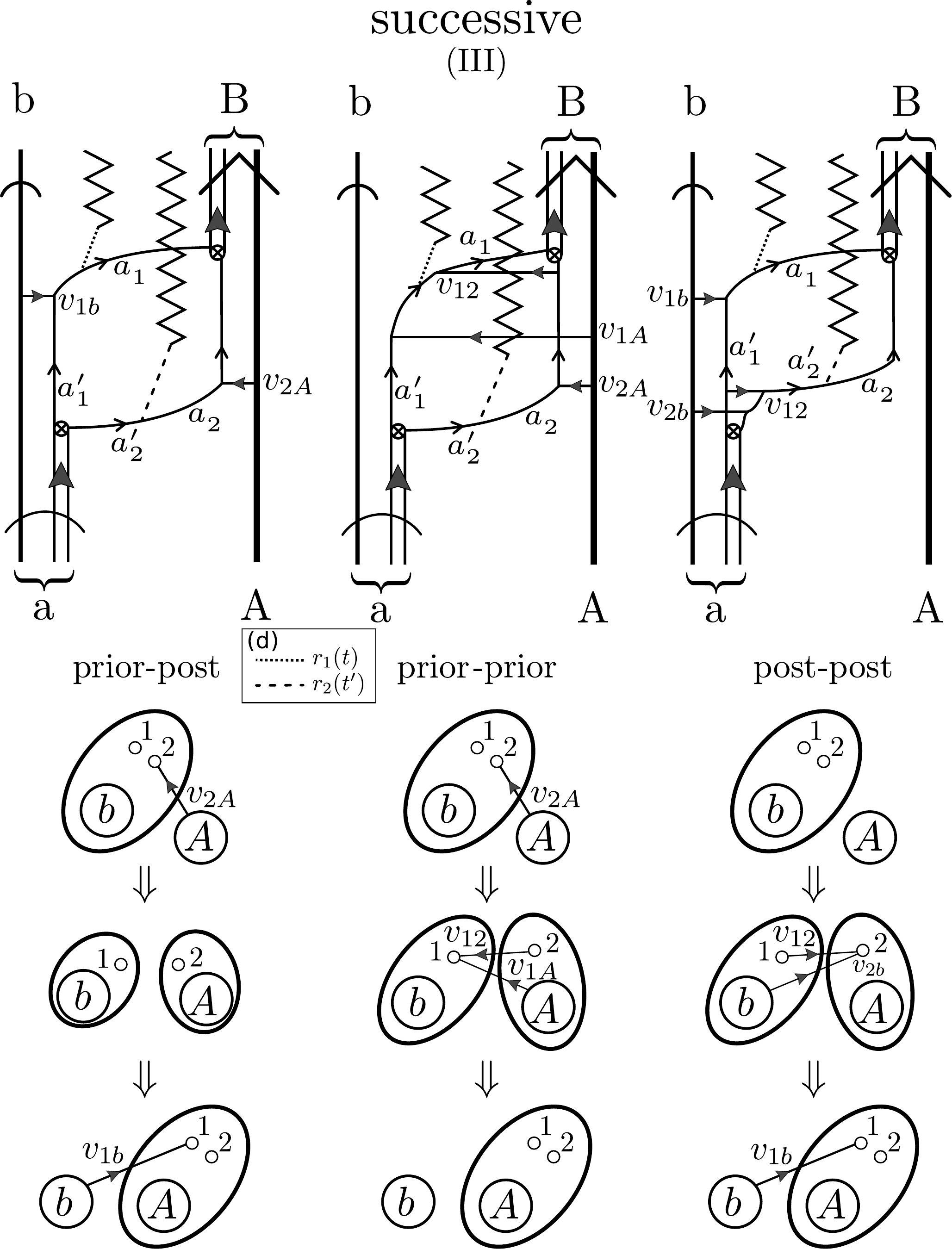}
	\end{center}
	\captionsetup{singlelinecheck=off,justification=raggedright}
	\caption{Graphical representation of the lowest order ((I),(II) and (III) first and second order in $v$ respectively), two-nucleon transfer processes, which correctly converge to the strong-correlation (only simultaneous transfer), and to the independent-particle (only successive transfer) limits. The time arrow is assumed to point upwards:
(I) Simultaneous transfer, in which one particle is transferred by the nucleon-nucleon interaction (note that $U(r)=\int d^3 r' \rho(r')v(|\vec r-\vec r'|)$ ) acting either in the entrance $\alpha \equiv a+A$ channel (prior) or in the final $\beta \equiv b + B$ channel (post), while the other particle follows suit making use of the particle-particle correlation (grey area) which binds the Cooper pair (see upper inset labelled (a)), represented by a solid arrow on a double line, to the projectile (curved arrowed lines) or to the target (opened arrowed lines). The above argument provides the explanation why when e.g. $v_{1b}$ acts on one nucleon, the other nucleon also reacts instantaneously. In fact a Cooper pair displays generalized rigidy (emergent property in gauge space).
A crossed open circle represents the particle-pair vibration coupling. The associated strength, together with an energy denominator, determines the amplitude $X_{a'_1 a'_2}$ (cf. Table 1) with which the pair mode (Cooper pair) is in the (time reversered) two particle configuration $a'_1 a'_2$. In the transfer process, the orbital of relative motion changes, the readjustement of the corresponding trajectory mismatch being operated by a Galilean operator ($\textrm{exp}\{ \vec k \cdot (\vec{r}_{1A}(t)+\vec{r}_{2A}(t))\}$). This phenomenon, known as recoil process, is represented by a jagged line which  provides simultaneous information on the two transferred nucleons (single time appearing as argument of both single-particle coordinates $r_1$ and $r_2$; see inset labeled (b)). In other words, information on the coupling of structure and reaction modes.
(II) Non-orthogonality contribution. While one of the nucleons of the Cooper pairs is transferred under the action of $v$, the other goes, uncorrelatedly over, profiting of the non-orthogonality of the associated single-particle wavefunctions (see inset (c)). In other words of the non-vanishing values of the overlaps, as shown in the inset.
(III) Successive transfer. In this case, there are two time dependences associated with the acting of the nucleon-nucleon interaction twice (see inset (d)).}
\label{fig:11}
\end{figure}

\begin{figure}[h!]
	\begin{center}
		\includegraphics[width=0.88\textwidth]{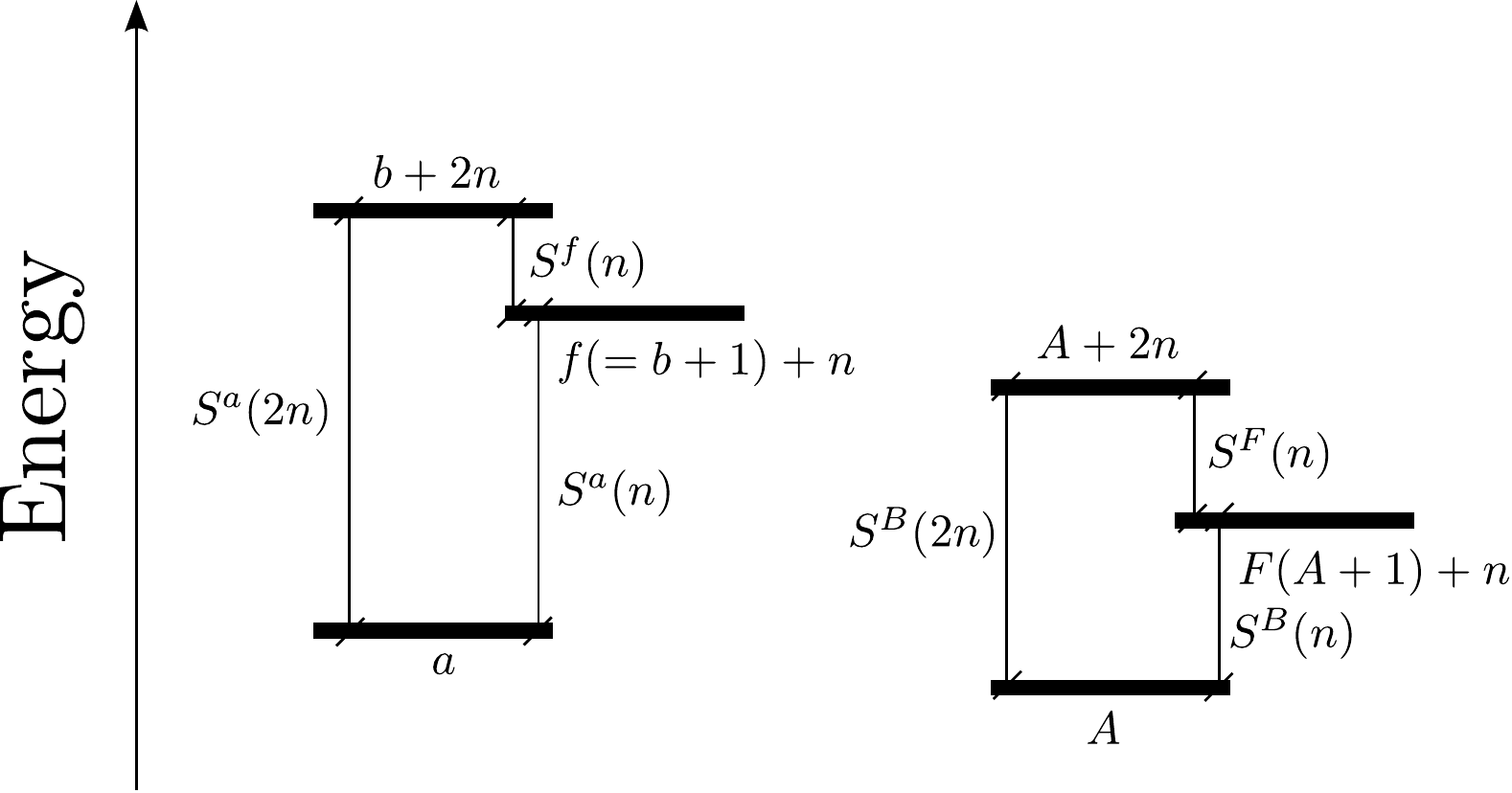}
	\end{center}
    \captionsetup{singlelinecheck=off,justification=raggedright}
	\caption{One--and two--neutron separation energies $S(n)$ and $S(2n)$ associated with	the channels\\ $\alpha \equiv a(b+2)+A \longrightarrow \gamma \equiv f (= b+1)+F (= A+1) \longrightarrow \beta \equiv b+B(= A+2)$.}
\label{fig:13a}
\end{figure}

\begin{figure}[h!]
	\begin{center}
		\includegraphics[width=0.88\textwidth]{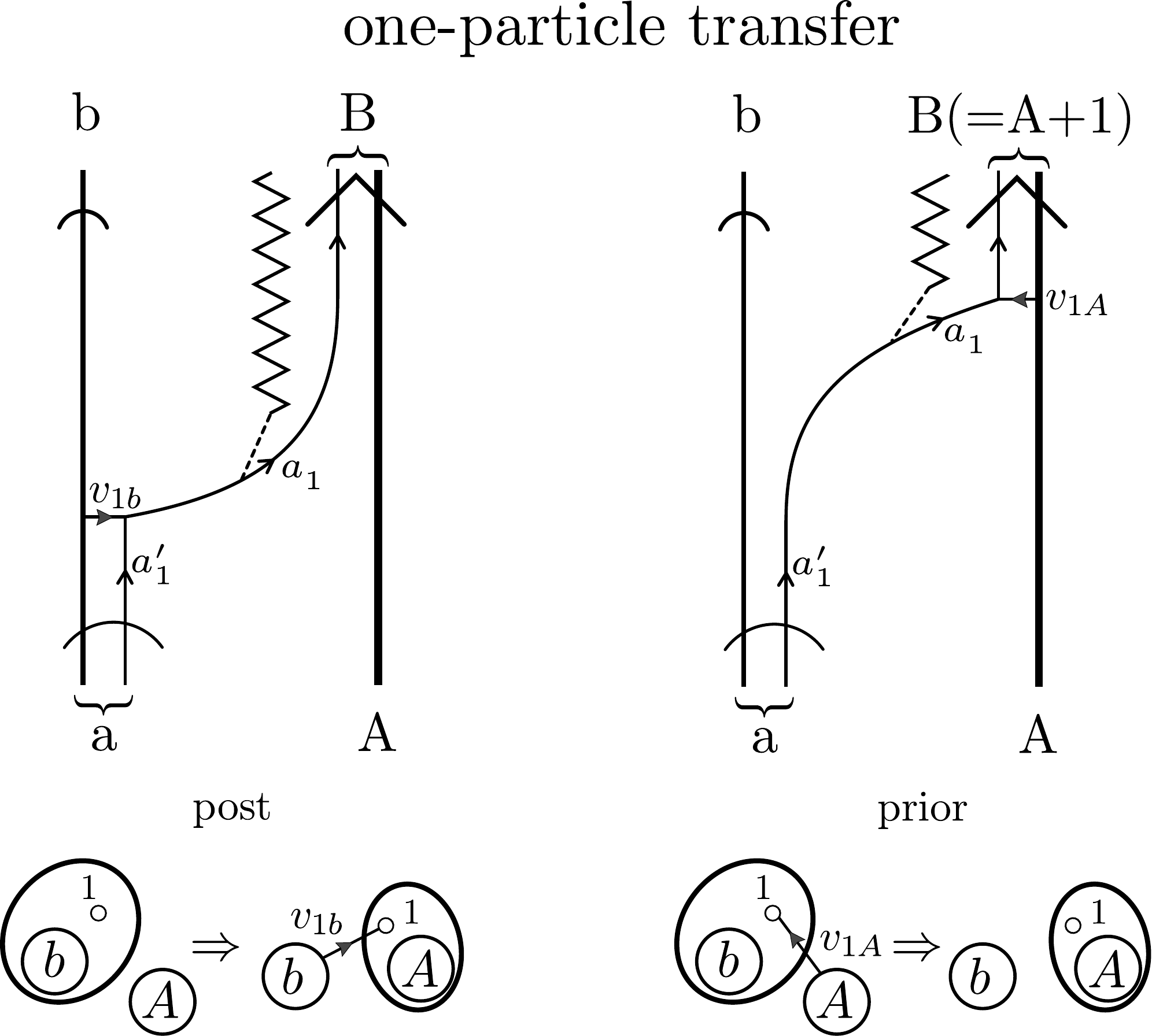}
	\end{center}
	\caption{One-particle transfer reactions $a(= b + 1) + A \longrightarrow b + B(= A + 1)$. The different symbols have been defined in the caption to Fig. 7.}
\label{fig:13b}
\end{figure}

\begin{figure}[h!]
	\begin{center}
		\includegraphics[width=\textwidth]{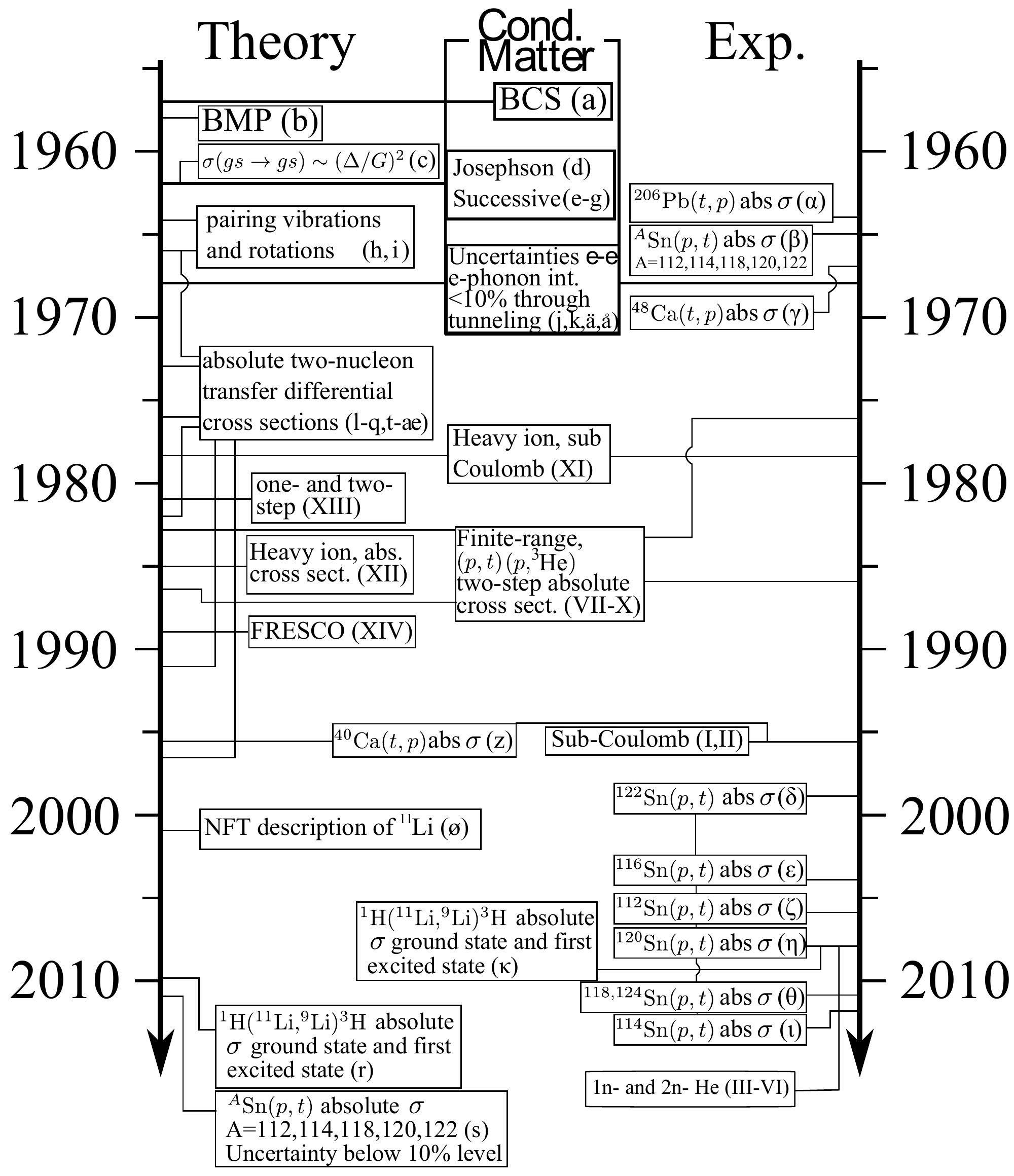}
	\end{center}
	\captionsetup{singlelinecheck=off,justification=raggedright}
	\caption{A schematic chronology of some of the milestones in the development of nuclear Cooper pair transfer as a quantitative tool to probe pairing correlations in nuclei (theory left, experiment right). In the center, the condensed matter reference events, starting from BCS theory and ending with a quantitative determination, through Cooper pair tunneling experiments, of the electron--electron and electron--phonon coupling formfactors.}
\label{fig:13c}
\end{figure}
\clearpage
 \textbf{Comments to Fig 10}. While it took the condensed matter community slightly over a decade from the publication of the BCS paper to make Cooper pair tunneling a quantitative tool to probe  pairing correlations in metals with uncertainties well below the $10\%$ level, half a century elapsed after the publication of the BMP paper (ref. (b)), which started the field of nuclear BCS, before the same level of accuracy was achieved with two-nucleon transfer reactions. A partial explanation of this fact can be found in the development which took place in the late seventies in which  nuclear reactions were tantamount to heavy ion reactions. 
 Such a situation  discouraged the continuation and improvement of two-nucleon transfer absolute differential cross sections to single, individual states, and the closing of essentially all of the $(t,p)$ facilities. One could furthermore argue that within the first community one found the likes of Bardeen, Cooper, Schrieffer, Josephson, Anderson, Giaver, Pines, Bogolyubov, Fr\"olich, Falicov, Cohen and Valatin, among others. This is true, as it is also true that among nuclear practitioners one counted Bohr, Mottelson, Brown, Wei\ss{}kopf, Feshbach, Nilsson, Winther, Thouless, Arima, Hansen and Nathan, among others.\\
 \\
 \\
	\noindent
	\textbf{(a)} J. Bardeen, L. N. Cooper, and J. R. Schrieffer, Theory of superconductivity, {\it Physical Review}. \textbf{106}, 162 (1957); ibid. \textbf{108}, 1175, (1957).\\
	\textbf{(b)} A. Bohr, B. R. Mottelson, and D. Pines, Possible analogy between the excitation spectra of nuclei and those of the superconducting metallic state, {\it Physical Review}. {\bf 110}, 936, (1958).\\
	\textbf{(c)} S. Yoshida, Note on the two-nucleon stripping reaction, {\it Nucl Phys.} {\bf 33}, 685, (1962). \\ %
	\textbf{(d)} B. D. Josephson, Possible new effects in superconductive tunnelling, {\it Phys. Lett.} {\bf 1}, 251, (1962). \\
	\textbf{(e)} J. Bardeen, Tunneling into superconductors, {\it Physical Review Letters}. \textbf{9}, 147, (1962). \\
	\textbf{(f)} M. H. Cohen, L. M. Falicov, and J. C. Phillips, Superconductive tunneling, {\it Phys. Rev. Lett.} {\bf 8}, 316, (1962). \\
	\textbf{(g)} P. W. Anderson, Special Effects in Superconductivity, {\it in The Many–-Body Problem}, Ed. E. R. Caianello Vol.2. (Academic Press, New York, 1964), pp. 113-135. \\
	\textbf{(h)} A. Bohr. In {\it Comptes Rendus du Congr\`{e}s International de Physique Nucl\'eaire, vol. 1, p. 487. Centre National de la Recherche Scientifique,} (1964). \\
	\textbf{(i)} D. R. B\`es and R. A. Broglia, Pairing vibrations, {\it Nucl. Phys.} {\bf 80}, 289, (1966). \\
	\textbf{(j)} D. J. Scalapino, The electron-phonon interaction and strong-coupling superconductors, pp. 449, {\it Superconductivity}, ed. R. Parks, (Marcel Dekker, New York, 1968) \\
	\textbf{(k)} W. L. McMillan and J. M. Rowell, Tunneling and strong-coupling superconductivity, pp. 561, {\it Superconductivity}, ed. R. Parks, (Marcel Dekker, New York, 1968) \\
	\textbf{(l)} R. A. Broglia, The pairing model, {\it Annals of Physics}. {\bf 80}, 60, (1973). \\
	\textbf{(m)} R. A. Broglia, O. Hansen, and C. Riedel, Two–neutron transfer reactions and the pairing model, {\it Adv. in Nucl. Phys.} {\bf 6}, 287, (1973). \url{http://merlino.mi.infn.it/repository/BrogliaHansenRiedel.pdf}. \\
	\textbf{(n)} T. Udagawa and D. Olsen, Note on the forbidden $(p, t)$ excitation of unnatural parity final states from $0^+$ targets via multistep processes, \textit{Phys. Lett. B}, {\bf 46}, 285, (1973). \\
	\textbf{(o)} M. Schneider, J. Burch, and P. Kunz, Competition of two-step processes in the reactions $^{60,62}$Ni$(p,t)$ leading to unnatural parity states, {\it Phys Lett B} {\bf63}, p. 129, (1976). \\
	\textbf{(p)} B. F. Bayman and J. Chen, One-step and two-step contributions to two-nucleon transfer reactions, {\it Phys. Rev. C} {\bf 26}, 1509, (1982).\\
	\textbf{(q)} W. S. Chien et al., Unnatural parity transitions in $^{22}$Ne$(p,t)^{20}$Ne, {\it Phys. Rev. C}, {\bf 12}, 332, (1975). \\
	\textbf{(r)} G. Potel, et al., Evidence for phonon mediated pairing interaction in the halo of the nucleus $^{11}$Li, {\it Phys. Rev. Lett.} {\bf 105}, 172502, (2010).\\
	\textbf{(s)} G. Potel, et al., Calculation of the Transition from Pairing Vibrational to Pairing Rotational Regimes between Magic Nuclei $^{100}\mathrm{Sn}$ and $^{132}\mathrm{Sn}$ via Two-Nucleon Transfer Reactions, {\it Phys. Rev. Lett.} {\bf 107}, 092501, (2011). \\
	\textbf{(t)} N. B. de Takacsy, Two-step mechanism in the reaction $^{208}$Pb$(p, t)$, {\it Phys. Rev. Lett.}, {\bf 31}, 1007,(1973). \\
	\textbf{(u)} N. B. de Takacsy, On the contribution from a two-step mechanism, involving the sequential transfer of two neutrons, to the calculation of $(p, t)$ reaction cross sections, \textit{Nucl. Phys.}, {\bf A 231}, 243, (1974). \\
	\textbf{(v)} N. Hashimoto and M. Kawai, The $(p, d)$ $(d, t)$ process in strong $(p, t)$ reactions, {\it Prog. Theor. Phys.}, {\bf 59},1245, (1978). \\
	\textbf{(w)} K. Kubo and H. Amakawa, Energy dependence of two-step $(p, t)$ cross sections, {\it Phys. Rev. C}, {\bf 17}, 1271, (1978). \\
	\textbf{(x)} K. Yagi et al., Anomalous analyzing powers for strong $(p_{pol} ,t)$ ground-state transitions and interference between direct and $(p,d)$ $(d,t)$ sequential process, {\it Phys. Rev. Lett.}, {\bf 43},1087, 1979. \\
	\textbf{(y)} M. Igarashi, K. Kubo, and K. Yagi, Two-nucleon transfer reaction mechanisms, {\it Phys. Rep.}, {\bf 199}, 1, (1991). \\
	\textbf{(z)} M. B. Becha et al., The $^{40}$Ca$(t,p)^{42}$Ca reaction at triton energies near 10 MeV per nucleon, {\it Phys. Rev. C}, {\bf 56}, 1960, (1997). \\
	\textbf{(\ae{})} H. Segawa, K. I. Kubo, and A. Arima, Two-step analysis of the $^{18}$O$(p, t)^{16}$O $2^-$ excitation and phase relations in the nonorthogonal term, {\it Phys. Rev. Lett.}, {\bf 35}, 357, (1975).\\
	\textbf{(\"{a})} I. Giaver, Electron Tunnelling and Superconductivity, \textit{Le Prix Nobel} 1973, Norstedt, Stockholm (1974) pp. 86-102.\\
	\textbf{(\aa{})} P. W. Anderson, Superconductivity in the Past and the Future, \textit{Superconductivity} Ed. R. Parks (Marcel Dekker, New York, 1968) pp. 1343-1358.\\
	\textbf{(\o{})} F. Barranco, et al., The halo of the exotic nucleus $^{11}${Li}: a single {C}ooper pair, {\it Europ. Phys. J. A} {\bf 11}, 385, (2001). \\
	\textbf{($\alpha$)} J. H. Bjerregaard, et al., States of $^{208}$Pb from double triton stripping, {\it Nucl. Phys.} {\bf 89}, 337, (1966). \\
	\textbf{($\beta$)} G. Bassani, et al., $(p,t)$ ground-state $L = 0$ transitions in the even isotopes of Sn and Cd at 40 MeV, $N = 62$ to 74,” {\it Phys. Rev.}, {\bf 139}, B830, (1965). \\
	\textbf{($\gamma$)} J. H. Bjerregaard, et al., The $(t,p)$ reaction with the even isotopes of Ca, {\it Nucl. Phys.} {\bf A 103}, 33, (1967). \\
	\textbf{($\delta$)} P. Guazzoni, et al., Level structure of $^{120}$Sn: High resolution $(p,t)$ reaction and shell model description, {\it Phys. Rev. C.} {\bf 60}, 054603, (1999).\\ 
	\textbf{($\varepsilon$)} P. Guazzoni, et al., High-resolution study of the $^{116}$Sn$(p,t)$ reaction and shell model structure of $^{114}$Sn,
	Phys. Rev. C. {\bf 69}, 024619, (2004). \\
	\textbf{($\zeta$)} P. Guazzoni, et al., Spectroscopy of $^{110}$Sn via the high-resolution $^{112}$Sn$(p,t)^{110}$Sn reaction, {\it Phys. Rev. C.} {\bf 74}, 054605, (2006). \\
	\textbf{($\eta$)} P. Guazzoni, et al., $^{118}$Sn levels studied by the $^{120}$Sn$(p,t)$ reaction: High-resolution measurements, shell model, and distorted-wave Born approximation calculations, {\it Phys. Rev. C.} {\bf 78}, 064608, (2008). \\
	\textbf{($\theta$)}  P. Guazzoni, et al., High-resolution measurement of the $^{118,124}$Sn$(p,t)^{116,122}$Sn reactions: Shell-model and microscopic distorted-wave Born approximation calculations, {\it Phys. Rev. C.} {\bf 83}, 044614, (2011). \\
	\textbf{($\iota$)} P. Guazzoni, et al., High resolution spectroscopy of $^{112}$Sn through the $^{114}$Sn$(p,t)^{112}$Sn reaction, {\it Phys. Rev. C} {\bf 85}, 054609, (2012). \\
	\textbf{($\kappa$)} I. Tanihata et al., Measurement of the two-halo neutron transfer reaction $^{1}$H($^{11}$Li,$^{9}$Li)$^{3}$H at 3A MeV, {\it Phys. Rev. Lett.} {\bf 100}, 192502, (2008). \\
        \textbf{(I)} L. Corradi et al., Single and pair neutron transfers at sub-barrier energies, {\it Phys. Rev. C} {\bf 84}, 034603 (2011). \\
        \textbf{(II)} P. A. DeYoung et al., Two-neutron transfer in the $^{6}$He+$^{209}$Bi reaction near the Coulomb barrier, {\it Phys. Rev. C} {\bf 71}, 051601 (2005). \\
        \textbf{(III)} A. Chatterjee et al., 1n and 2n transfer with the Borromean nucleus $^{6}$He near the Coulomb barrier, {\it Phys. Rev. Lett.} {\bf 101}, 032701 (2008). \\
        \textbf{(IV)} A. Lemasson et al., Reactions with the double-Borromean nucleus $^{8}$He, {\it Phys. Rev. C} {\bf 82}, 044617 (2010). \\
        \textbf{(V)} A. Lemasson et al., Pair and single neutron transfer with Borromean $^{8}$He, {\it Phys. Lett. B} {\bf 697}, 454 (2011). \\
        \textbf{(VI)} N. Keeley et al., Probing the $^{8}$He ground state via the $^{8}$He$(p,t)^6$He reaction, {\it Phys. Lett. B} {\bf 646}, 222 (2007). \\
        \textbf{(VII)} L. A. Charlton, Finite-range evaluation of $(p-d,d-t)$ with momentum space techniques, {\it Phys. Rev. C} {\bf 14}, 506 (1976). \\
        \textbf{(VIII)} M. Yasue et al., $^{12}$C$(p,t)^{10}$C and $^{12}$C$(p,^{3}$He$)^{10}$B reactions at $E_p = 51.9$ MeV, {\it Journ. of the Phys. Soc. of Jap.} {\bf 42}, 367 (1977). \\
        \textbf{(IX)} T. Takemasa, T. Tamura and T. Udagawa, Exact finite range calculations of light-ion induced two-neutron transfer reactions, {\it Nucl. Phys. A} {\bf 321}, 269 (1979) \\
        \textbf{(X)} K. Kurokawa et al., One- and two-step processes in natural and unnatural-parity $^{208}$Pb$(p,t)^{206}$Pb reaction at $E_p=22$ MeV, {\it Nucl. Phys. A} {\bf 470}, 377 (1987) \\
        \textbf{(XI)} M. A. Franey et al., Two neutron transfer reaction mechanism with heavy ions at sub-Coulomb energies, {\it Phys. Rev. Lett.} {\bf 41}, 837 (1978). \\
        \textbf{(XII)} E. Maglione et al., Absolute cross sections of two-nucleon transfer reactions induced by heavy ions, {\it Phys. Lett. B} {\bf 162}, 59 (1985). \\
        \textbf{(XIII)} B. F. Bayman and J. Chen, One-step and two-step contributions to two-nucleon transfer reactions, {\it Phys. Rev. C} {\bf 26}, 1509 (1982). \\
        \textbf{(XIV)} I. J. Thompson, Coupled reaction channels calculations in nuclear physics, {\it Computer Phys. Rep.}, {\bf 7}, 167 (1988).
\clearpage 
\bibliographystyle{ieeetr} 


\end{document}